\documentclass[10pt,conference]{IEEEtran}
\usepackage{etoolbox}
\usepackage{booktabs} 

\usepackage{balance}
\usepackage{graphicx}
\usepackage{subcaption}
\usepackage{multirow}
\usepackage{listings}
\usepackage{fancybox}
\usepackage{graphicx}
\usepackage[table]{xcolor}
\usepackage{algorithm}
\usepackage{algorithmic}
\usepackage{amssymb}
\usepackage{pgfplots}
\usepackage{cite}
\makeatletter
\newcommand{\mybox}[1]{%
  \setbox0=\hbox{#1}%
  \setlength{\@tempdima}{\dimexpr\wd0+13pt}%
  \begin{tcolorbox}[boxrule=0.5pt, colback=white, arc=4pt,
	        left=6pt,right=6pt,top=6pt,bottom=6pt,boxsep=0pt]
			    #1
  \end{tcolorbox}
}

\definecolor{mycolorhigh}{RGB}{168, 157, 89}
\definecolor{mycolormiddle}{RGB}{224, 224, 182}

\definecolor{mygray}{rgb}{0.61, 0.61, 0.61}

\definecolor{mypink}{rgb}{0.96, 0.76, 0.76}

\newcommand{\task}{218}
\newcommand{\company}{BigCompany}
\newcommand{\tool}{\textsc{iSENSE}}

\ifCLASSINFOpdf
\else
\fi
\begin{document}
%
\title{Effective Automated Decision Support for Managing Crowdtesting}

\author{\IEEEauthorblockN{Junjie Wang$^1$, Ye Yang$^2$, Rahul Krishna$^3$, Tim Menzies$^3$, Qing Wang$^1$}
\IEEEauthorblockA{$^1$Institute of Software Chinese Academy of Sciences, Beijing, China\\
$^2$School of Systems and Enterprises, Stevens Institute of Technology, USA\\
$^3$Department of Computer Science, North Carolina State University, Raleigh, NC, USA\\
Email: {wangjunjie,wq}@itechs.iscas.ac.cn, ye.yang@stevens.edu, zyu9@ncsu.edu, tim@menzies.us}
}


%


\maketitle

\begin{abstract}


Crowdtesting has grown to be an effective alternative to traditional testing, especially in mobile apps. However, crowdtesting is hard to manage in nature. Given the complexity of mobile applications and unpredictability of distributed, parallel crowdtesting process, it is difficult  to  estimate  (a)  the  remaining number of bugs as yet undetected or (b) the required cost to find those  bugs.  Experience-based  decisions  may  result  in  ineffective crowdtesting  process.

This paper aims at exploring automated decision support to effectively manage crowdtesting process. The proposed {\tool} applies incremental sampling technique to process crowdtesting reports arriving in chronological order, organizes them into fixed-size groups as dynamic inputs, and predicts two test completion indicators in an incrementally manner. 
The two indicators are: 1) total number of bugs predicted with Capture-ReCapture (CRC) model, and 2) required test cost for achieving certain test objectives
predicted with AutoRegressive Integrated Moving Average
(ARIMA) model.
We assess {\tool} using  46,434 reports of 218 crowdtesting tasks from one of the largest crowdtesting platforms in China. 
Its effectiveness is demonstrated through two applications for automating crowdtesting management, i.e. automation of task closing decision,
and semi-automation of task closing trade-off analysis.
The results show that decision automation using {\tool} will provide managers with greater opportunities to achieve cost-effectiveness gains of crowdtesting. 
Specifically, a median of 100\% bugs can be detected with 30\% saved cost based on the automated close prediction.

\end{abstract}


%
\IEEEpeerreviewmaketitle

\section{Introduction}
\label{sec:intro}

Crowdtesting is an emerging trend in software testing which accelerates testing process by attracting online crowdworkers to accomplish various types of testing tasks \cite{mao2017survey,chen2018research,link_crowdtest,wang2017domain,cui2017who}, esp. in mobile application testing.
It entrusts testing tasks to crowdworkers whose diverse testing environments/platforms, background, and skill sets could significantly contribute to more reliable, cost-effective, and efficient testing results.

Trade-offs such as ``how much testing is enough'' are critical yet challenging project decisions in software engineering \cite{myers2011art,lewis2016software,garg2011stop,iqbal2013software}.
Insufficient testing can lead to unsatisfying software quality, while excessive testing can result in potential schedule delays and low cost-effectiveness. 

Many existing approaches employed either risk-driven or value-based analysis for prioritizing or selecting test cases, and minimizing test runs \cite{wang2017qtep,shi2015comparing,harman2015empirical,saha2015information,henard2016comparing}, in order to effectively plan and manage testing process. 
However, none of these is applicable to the emerging crowd testing paradigm where managers typically have no control over online crowdworkers' dynamic behavior and uncertain performance. 
Worse still, there is no existing method to support crowdtesting management.

Consequently, due to lack of decision support, in practice, project managers typically plan for the close time of crowdtesting tasks solely based on their personal experience.
However, it is very challenging for  managers to come up with reasonable experience-based crowdtesting decisions. 
This is because our investigation on real-world crowdtesting data (Section \ref{subsec:background_observations}) reveals that there are large variations in bug arrival pattern of crowdtesting tasks, and in task's duration and consumed cost for achieving the same quality level.  
Furthermore, crowdtesting is typically treated as black box and managers' decisions remain insensitive to actual testing progress. 
Hence, managers have significant challenges deciding when to intervene and close the task (and improve cost-effectiveness as well). 


To address these challenges, this paper aims at exploring automated decision support to effectively manage crowdtesting. 
In detail, we focus on exploring dynamical bug arrival data associated with crowdtesting reports, and investigate whether it is possible to determine that, at certain point of time, a task has achieved satisfactory bug detection level (e.g. indicated by a percentage), based on the dynamic bug arrival data.

The proposed {\tool}\footnote{{\tool} is named considering it likes a sensor in crowdtesting process to raise the awareness of the testing progress.} applies incremental sampling technique to process crowdtesting reports arriving in chronological order,  organizes them into fixed-size groups as dynamic inputs, and integrates the Capture-ReCapture (CRC) model and the Autoregressive Integrated Moving Average (ARIMA) model to raise awareness of crowdtesting progress.  
CRC model is widely applied to estimate the total population based on the overlap generated by multiple captures \cite{rong2017towards,liu2015adoption,chun2006estimating,mandala2012application}. 
ARIMA model is commonly used to model time series data to forecast the future trend \cite{amin2013approach,chong1988analyzing,kemerer1999empirical,herraiz2007forecasting}.
{\tool} predicts two test completion indicators in an incrementally manner, including: 1) total number of bugs predicted with CRC model, and 2) required test cost for achieving certain test objectives predicted with ARIMA model. 
To the best of our knowledge, this is the first study to apply incremental sampling technique in crowdtesting management, so as to better model the bug arrival dynamics.

{\tool} is evaluated using {\task} crowdtesting tasks from one of the largest Chinese crowdtesting platforms. 
Results show that, the median errors on {\tool}'s  prediction performance (of total bugs, and required cost) are both below 3\%, with about 10\% standard deviation during the second half of the crowdtesting process.

We further demonstrate its applications through two typical decision scenarios, one for automating task closing decision, and the other for semi-automation of task closing trade-off analysis. 
The results show that decision automation using {\tool} will provide managers with greater opportunities to achieve cost-effectiveness gains of crowdtesting. 
Specifically, a median of 100\% bugs can be detected with 30\% saved cost based on  the  automated  close  prediction.

The contributions of this paper are as follows:

\begin{itemize}

\item Empirical observations on crowdtesting bug arrival patterns based on industrial dataset, which has motivated this study and can motivate future studies.

\item Integration of incremental sampling technique to model crowdtesting bug arrival data.

\item Development of CRC-based model for predicting total number of bugs, and ARIMA-based model for predicting required cost for achieving certain test objectives. 

\item {\tool} approach for automated decision support in crowdtesting management, including automating task closing decision, and semi-automation of task closing trade-off analysis.

\item Evaluation of {\tool} on 46,434 reports of {\task} crowdtesting tasks from one of the largest crowdtesting platforms in China, and results are promising\footnote{Url for {\tool} website with experimental dataset, source code and detailed experimental results is blinded for review}.



\end{itemize}

\section{Background and Motivation}
\label{sec:background}

\subsection{Background}
\label{subsec:background_crowdtesting}

In general crowdtesting practice, managers prepare the crowtesting task (including the software under test and test requirements), and distribute it on certain online crowdtesting platform. Crowdworkers can sign in their interested tasks and submit test reports, typically summarizing test input, test steps, test results, etc.

The crowdtesting platform receives and manages crowdtesting reports submitted by the crowdworkers. Project managers then inspect and verify each report for their tasks manually or using automatic tool support (e.g., \cite{wang2016towards,wang2017domain} for automatic report labeling). 
Generally, each report will be characterized using two attributes: 1) whether it contains a valid bug\footnote{In our experimental platform, a report corresponds to either 0 or 1 bug, and there is no reports containing more than 1 bugs.}; 2) if yes, whether it is a duplicate bug that has been previously reported by other crowdworkers. 

In the following paper, if not specified, when we say ``\textit{bug}'' or ``\textit{unique bug}'', we mean the corresponding report contains a bug and the bug is not the duplicate of previously submitted ones. 

\begin{figure*}[t!]
  \centering
 \hspace{-0.2in}
   \begin{subfigure}{0.24\textwidth}
    \includegraphics[width=4.3cm]{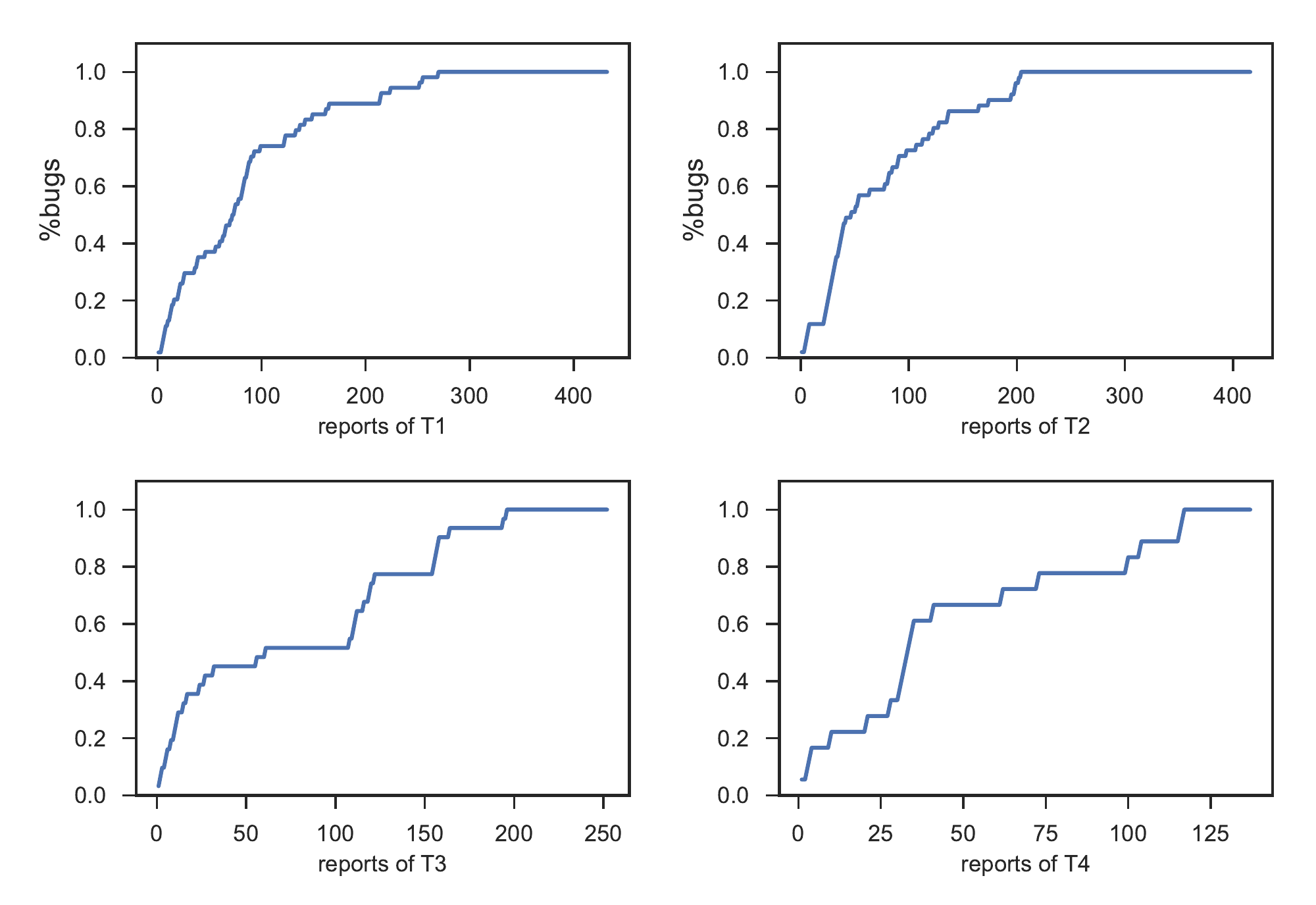}
   \caption{Example bug arrival curve}
   \label{fig:bugRateCurve}
  \end{subfigure}
    \hspace{-0.1in}
  \begin{subfigure}{0.24\textwidth}
    \includegraphics[width=4.3cm]{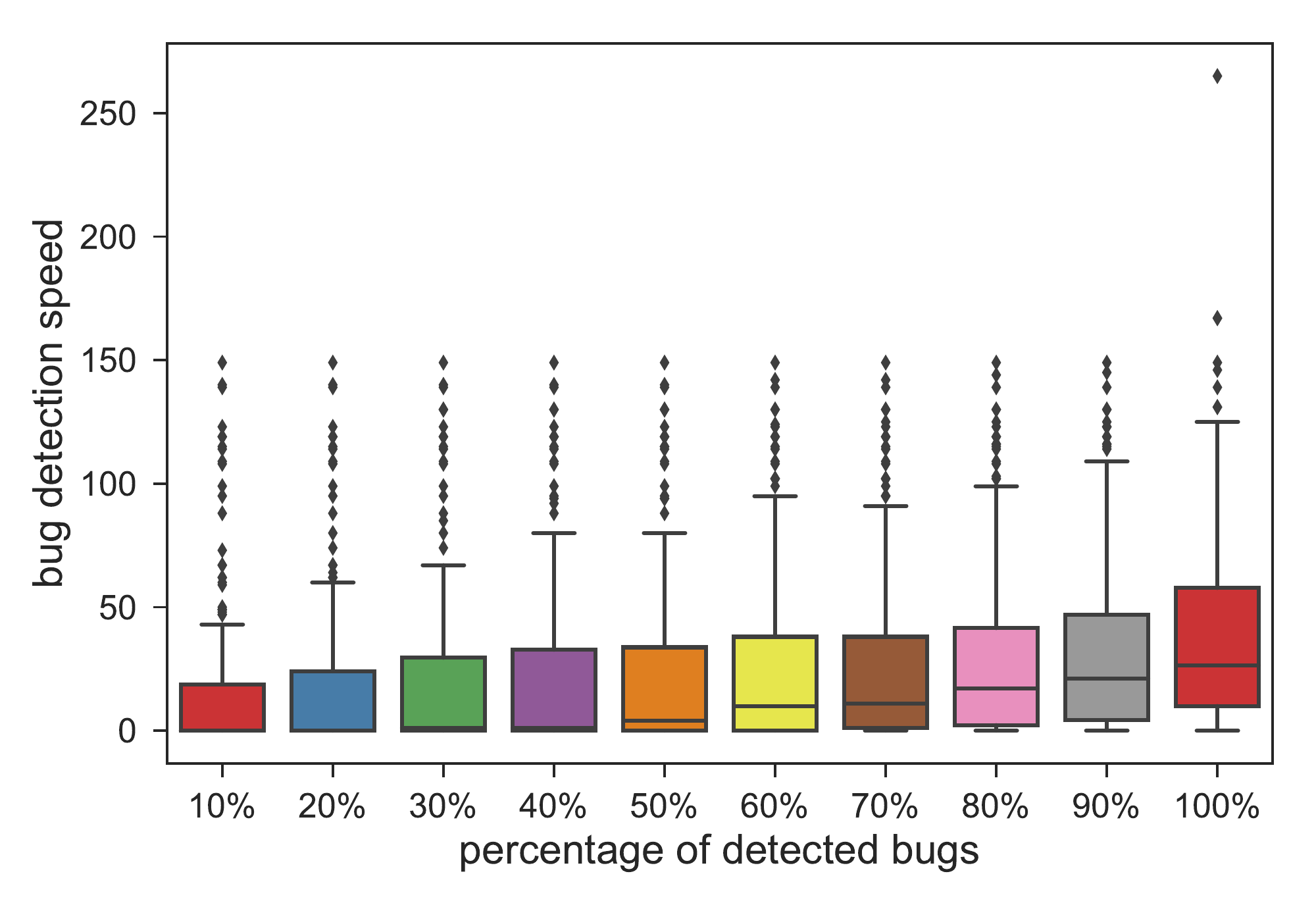}
   \caption{Bug detection speed}
   \label{fig:duration}
  \end{subfigure}
 \hspace{-0.1in}
  \begin{subfigure}{0.24\textwidth}
    \includegraphics[width=4.3cm]{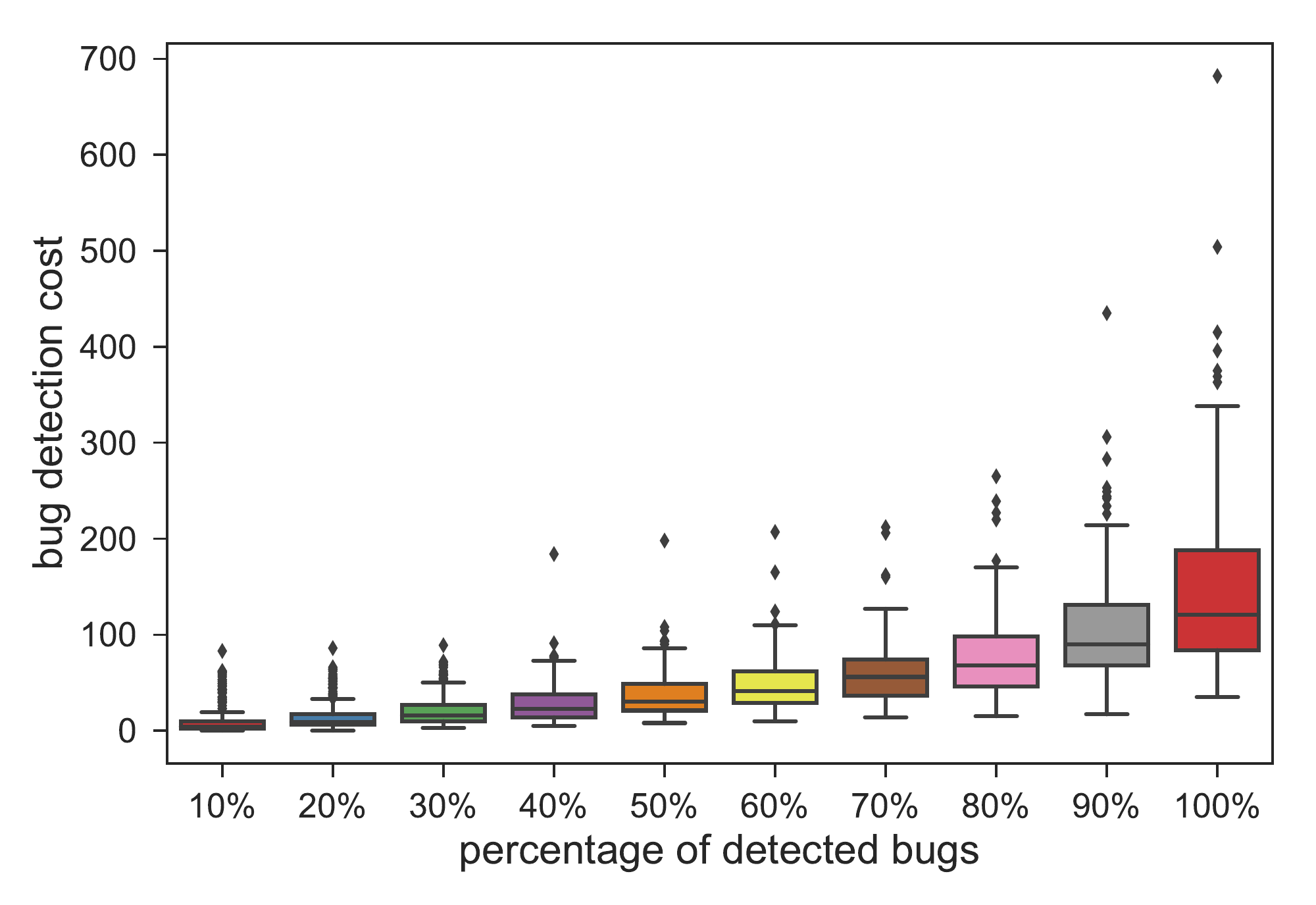}
  \caption{Bug detection cost}
   \label{fig:cost}
  \end{subfigure}
  \begin{subfigure}{0.24\textwidth}
    \includegraphics[width=4.3cm]{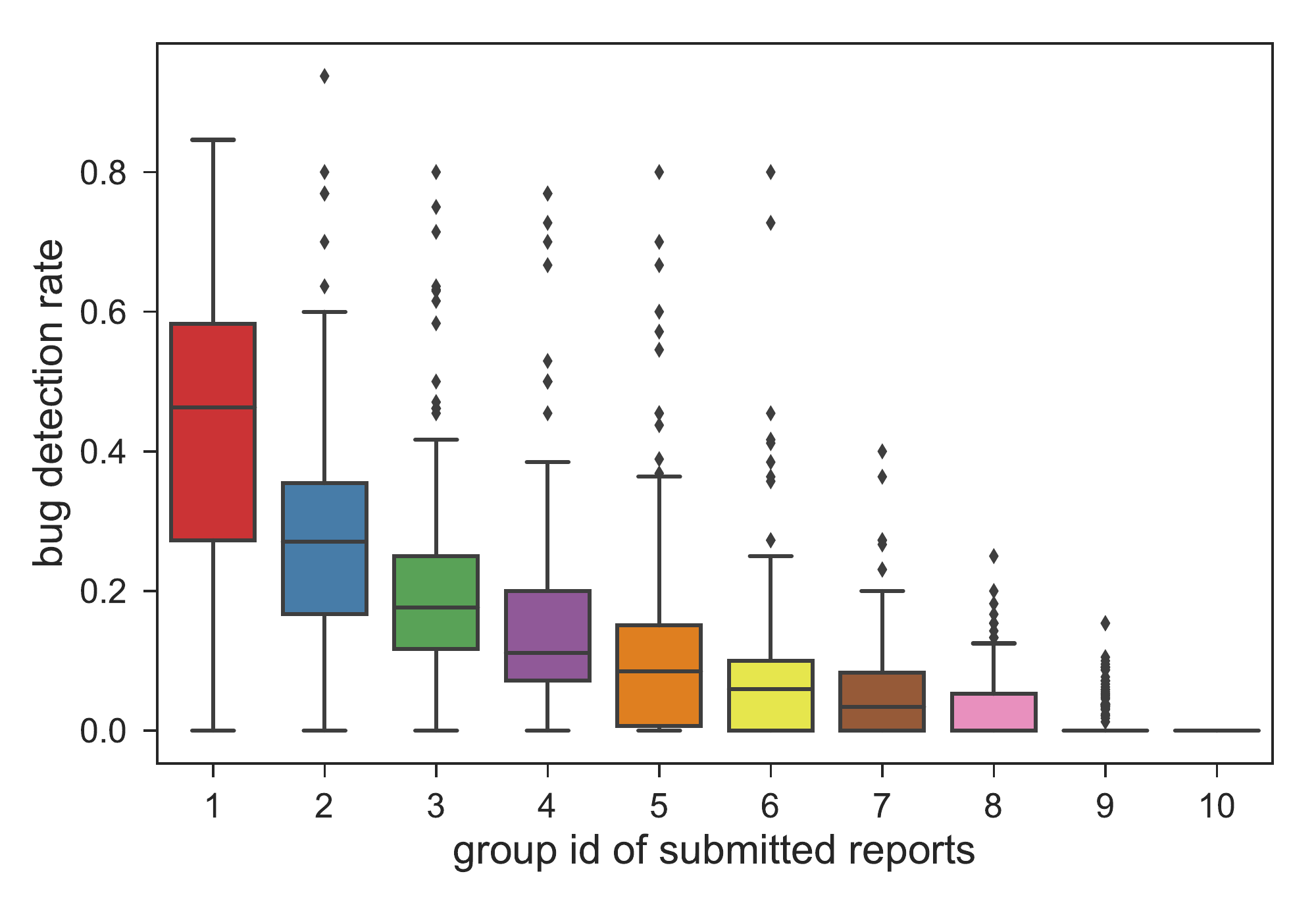}
  \caption{Bug detection rate}
   \label{fig:bugRate}
  \end{subfigure}
  \caption{Observations on real-world crowdtesting data}
  \label{fig:observation}
\end{figure*}%
\subsection{{\company} DataSet}
\label{subsec:background_dataset}

Our experimental dataset is collected from {\company}\footnote{Blinded for review.} crowdtesting platform, which is one of the largest  platforms in China. The dataset contains all tasks completed between May. 1st 2017 and Jul. 1st 2017.
In total, there are 218 tasks, with 46434 submitted reports.
The minimum, average, and maximum number of crowdtesting reports (\textit{and unique bugs}) per task are  101 (\textit{6}), 213 (\textit{26}), and 876 (\textit{89}), respectively.

\subsection{Observations From A Pilot Study}
\label{subsec:background_observations}

To understand the bug arrival patterns of crowdtesting, we conduct a pilot study to analyze three bug detection metrics, i.e. \textit{bug detection speed}, \textit{bug detection cost}, and \textit{bug detection rate}. 

For each task, we first identify the time when K\% bugs have been detected, 
where we treat the number of historical detected bugs as the total number. 
K is ranged from 10 to 100.
Then, the \textit{bug detection speed} for a task can be derived using the duration (measured in \textit{hours}) between its open time and the time it receives K\% bugs. 
Next, the \textit{bug detection cost} for a task can be derived using the number of submitted reports\footnote{Note that, the primary cost in crowdtesting is the reward to crowdworkers,  and their submitted reports are usually equally paid \cite{chen2018research,cui2017who}; Hence, the number of received reports is treated as the consumed cost for simplicity in this study.} by reaching K\% bugs.

To examine \textit{bug detection rate}, we break the crowdtesting reports for each task into 10 equal-sized groups, in chronological order. The rate for each group is derived using the ratio between the number of unique bugs and the number of reports in the corresponding group. 

In addition, for each crowdtesting task, we also count the percentage of accumulated bugs (denoted as bug arrival curve) for the previous K reports, where K ranges from 1 to the total number of reports. 

Next, we present two general bug arrival patterns derived from the pilot study.

\subsubsection{\textbf{Large Variation in Bug Arrival Speed and Cost}}
\label{subsubsec:background_obs_durCost}

In Figure \ref{fig:bugRateCurve}, we first present four example bug arrival curves randomly selected from {\task} crowdtesting tasks, illustrating the diversity of bug arrival curves among different tasks.

In general, there is large variation in bug arrival speed and cost. 
Figure \ref{fig:duration} and \ref{fig:cost} demonstrates the distribution of bug detection speed and bug detection cost for all tasks.
It is obvious that, to achieve the same K\% bugs, there is large variation in both metrics. This is particularly true for a larger K\%.
For example, when detecting 90\% bugs, the bug detection speed ranges from 3 hours to 149 hours, while the bug detection cost ranges from 27 to 435 reports.

\subsubsection{\textbf{Decreasing Bug Arrival Rates Over Time}}
\label{subsubsec:background_obs_costEffi}

Figure \ref{fig:bugRate} shows the bug detection rate of the 10 break-down groups across all tasks.
We can see that the bug detection rate decreases sharply during the crowdtesting process.
This signifies that the cost-effectiveness of crowdtesting is dramatically decreasing for the later part of the process. 

In addition, from Figure \ref{fig:bugRateCurve}, we can also see that during the later part of the crowdtesting task, there is usually a flat area in bug arrival curve, denoting no new bugs submitted. 
This further suggests the potential opportunity for introducing automated closing decision support to increase cost-effectiveness of crowdtesting.

\subsubsection{\textbf{Needs of Automated Decision Support}}
\label{subsubsec:background_obs_demand}

In addition, an unstructured interview\footnote{We present the details about this interview in {\tool} website.} was conducted with the managers of {\company}, with findings shown below.

Project managers commented the black-box nature of crowdtesting process. While they can receive constantly arriving reports, they are often out of clue about the remaining number of bugs as yet undetected, or the required cost to find those additional bugs. 

Because they could not know what is going on of the crowdtesting, the management of crowdtesting is conducted as a guesswork.
This frequently results in many blind decisions in task planning and management.

\textbf{In Summary}, because there are large variations in bug arrival speed and cost (Section \ref{subsubsec:background_obs_durCost}), current decision making is largely done by guesswork. 
This results in low cost-effectiveness of crowdtesting (Section \ref{subsubsec:background_obs_costEffi}). 
A more effective alternative to manage crowdtesting would be to dynamically monitor the crowdtesting reports and automatically alert managers or close tasks when certain pre-specified test objectives are met, e.g. 90\% bugs have been detected, to save unnecessary cost wasting on later arriving reports.

Furthermore, current practice suggests a practical need to empower managers with greater visibility into the crowdtesting processes (Section \ref{subsubsec:background_obs_demand}), and ideally raise their awareness about task progress (i.e., remaining number of bugs, and required cost to meet certain test objectives), thus facilitate their decision making.

This paper intends to address these practical challenges by developing a novel approach for automated decision support in crowdtesting management, so as to improve cost-effectiveness of crowdtesting.

\section{Approach}
\label{sec:approach}

\begin{figure}[t!]
\centering
\includegraphics[width=7cm]{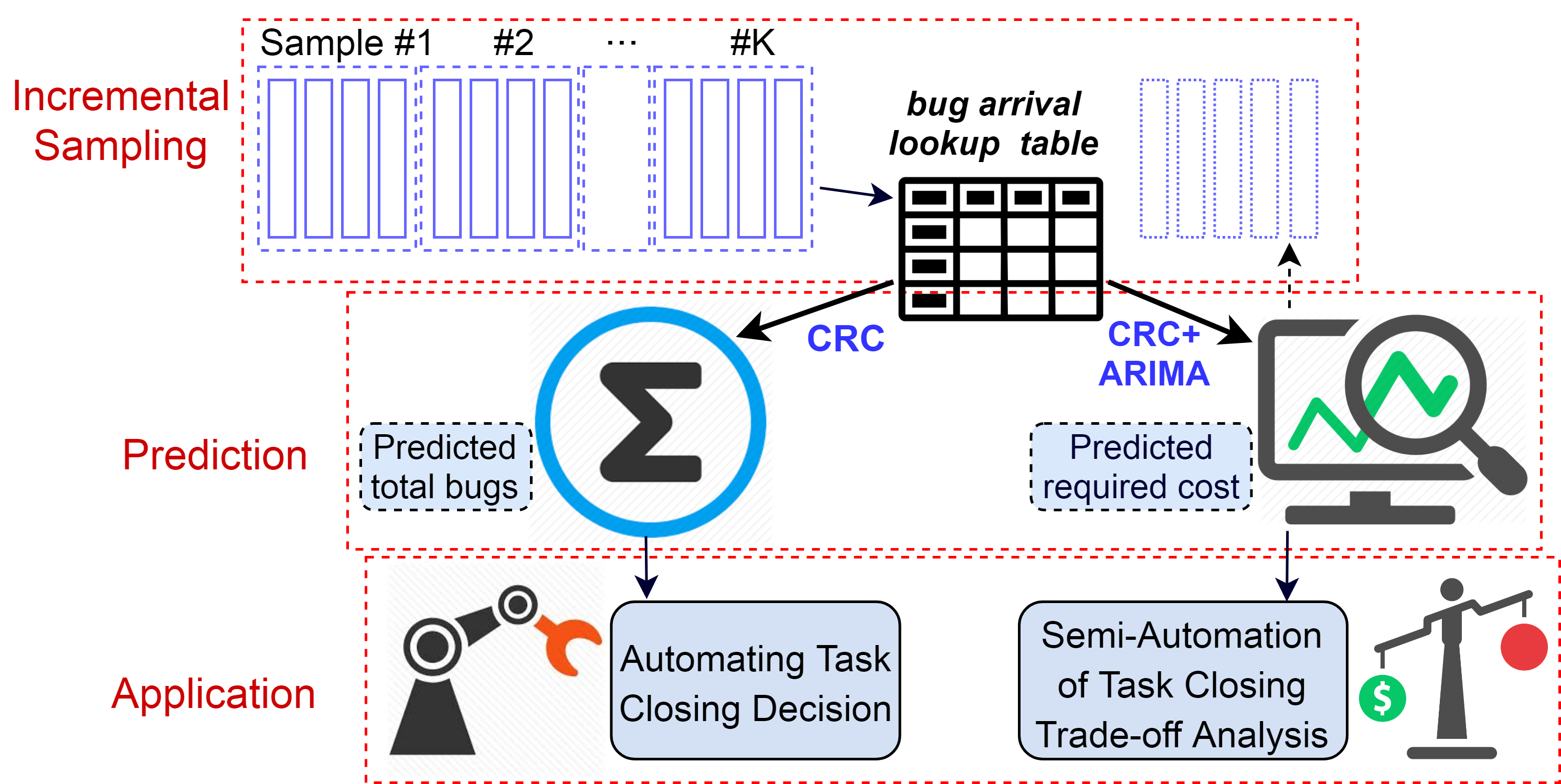}
\caption{Overview of {\tool}}
\label{fig:overview}
\end{figure}

Figure \ref{fig:overview} presents an overview of {\tool}. It consists of three main steps. First, {\tool} adopts an incremental sampling process to model crowdtesting reports. 
During the process, {\tool} converts the raw crowdtesting reports arrived chronologically into groups and generates a \textit{bug arrival lookup table} to characterize information on bug arrival speed and diversity. Then, {\tool}  integrates two models, i.e. CRC and ARIMA, to predict the total number of bugs contained in the software, and the required cost for achieving certain test objectives, respectively. Finally, {\tool} applies such estimates to support two typical crowdtesting decision scenarios, i.e., automating task closing decision, and semi-automation of task closing trade-off analysis. 
We will present each of the above steps in more details.

\subsection{Preprocess Data based on Incremental Sampling Technique}
\label{subsec:approach_preprocess}

Incremental sampling technique \cite{incrementalsampling} is a composite sampling and processing protocol. 
Its objective is to obtain a single sample for analysis that has an analytic concentration representative of the decision unit.
It improves the reliability and defensibility of sampling data by reducing variability when compared to conventional discrete sampling strategies.

Considering the submitted crowdtesting reports of chronological order (Section \ref{subsec:background_crowdtesting}), 
when \textit{smpSize} (\textit{smpSize} is an input parameter) reports are received, {\tool} treats it as a representative group to reflect the multiple parallel crowdtesting sessions. 
Remember in Section \ref{subsec:background_crowdtesting}, we mentioned that, each report is characterized as: 1) whether it contains a bug; 2) whether it is duplicate of previously submitted reports; if no, it is marked with a new tag; if yes, it is marked with the same tag as the duplicates. 
During the crowdtesting process, we maintain a two-dimensional \textit{Bug Arrival Lookup Table} to record these information (as Table \ref{tab:CRC-demo}).

\begin{table}[!h]
\caption{Example of bug arrival lookup table}
\label{tab:CRC-demo}
\centering
\scriptsize
\scalebox{0.97}{
\begin{tabular}{p{1.1cm}|p{0.1cm}|p{0.1cm}|p{0.1cm}|p{0.1cm}|p{0.1cm}|p{0.1cm}|p{0.1cm}|p{0.1cm}|p{0.1cm}|p{0.18cm}|p{0.15cm}|p{0.18cm}|p{0.15cm}}
\hline
& \textbf{\#1} & \textbf{\#2} & \textbf{\#3} & \textbf{\#4} & \textbf{\#5} & \textbf{\#6} & \textbf{\#7} & \textbf{\#8} & \textbf{\#9} & \textbf{\#10} & \textbf{\#11} & \textbf{\#12} & ...  \\
\hline
Sample \#1 & \cellcolor{mypink} 1 & \cellcolor{mypink} 1 & \cellcolor{mypink} 1 & 0 & 0 & 0 & 0 & 0 & 0 & 0 & 0 & 0  \\
\hline
Sample \#2 & 0 & 0 & \cellcolor{mypink} 1 & \cellcolor{mypink} 1 & 0 & 0 & 0 & 0 & 0 & 0 & 0 & 0  \\
\hline
Sample \#3 & 0 & 0 & \cellcolor{mypink} 1 & 0 & \cellcolor{mypink} 1 & 0 & 0 & 0 & 0 & 0 & 0 & 0  \\
\hline
Sample \#4 & 0 & 0 & 0 & \cellcolor{mypink} 1 & \cellcolor{mypink} 1 & \cellcolor{mypink} 1 & \cellcolor{mypink} 1 & \cellcolor{mypink} 1 & 0 & 0 & 0 & 0  \\
\hline
Sample \#5 & 0 & 0 & \cellcolor{mypink} 1 & \cellcolor{mypink} 1 & 0 & 0 & 0 & \cellcolor{mypink} 1 & \cellcolor{mypink} 1 & \cellcolor{mypink} 1 & \cellcolor{mypink} 1 & 0  \\
\hline
Sample \#6 & \cellcolor{mypink} 1 & 0 & \cellcolor{mypink} 1 & 0 & \cellcolor{mypink} 1 & 0 & 0 & 0 & 0 & 0 & 0 & \cellcolor{mypink} 1  \\
\hline
Sample \#7 & \multicolumn{12}{c}{...}  \\
\hline
\end{tabular}
}
\end{table}

After each sample is received, we first add a new row (suppose it is row \textit{i}) in the lookup table. 
We then go through each report contained in this sample.
For the reports not containing a bug, we ignore it.
Otherwise, if it is marked with the same tag as existing unique bugs (suppose it is column \textit{j}), record \textit{1} in row \textit{i} column \textit{j}.
If it is marked with a new tag, add a new column in the lookup table (suppose it is column \textit{k}), and record \textit{1} in row \textit{i} column \textit{k}.
For the empty cells in row \textit{i}, fill it with 0.

\subsection{Predict Total Bugs Using CRC}
\label{subsec:approach_bug_predict}

\subsubsection{\textbf{Background about CRC}}
\label{subsec:background_related_CRC}

CRC model, which uses the overlap generated by  multiple  captures  to  estimate  the  total  population,  has  been applied  in  software  inspections  to  estimate  the  total  number  of bugs \cite{rong2017towards,chun2006estimating,liu2015adoption,mandala2012application}. 
Existing CRC models can be categorized into four types according to bug detection probability (i.e. identical vs. different) and crowdworker's detection capability (i.e. identical vs. different), as shown in Table \ref{tab:CRC}.

Model \textit{M0} supposes all different bugs and crowdworkers have the same detection probability. Model \textit{Mh} supposes that the bugs have different probabilities of being detected. Model \textit{Mt} supposes that the crowdworkers have different detection capabilities. Model \textit{Mth} supposes different detection probabilities for different bugs and crowdworkers.

\begin{table}[!h]
\caption{Capture-ReCapture models}
\label{tab:CRC}
\centering
\scalebox{0.95}{
\scriptsize
\begin{tabular}{p{1.95cm}p{1cm}|p{2.0cm}p{2.0cm}}
 &  & \multicolumn{2}{c}{\bfseries{Crowdworker's detection capability}} \\
 & & Identical & Different \\
  \hline
\bfseries{Bug detection} & Identical  & M0 (\textit{M0})  & Mt (\textit{MtCH})  \\
\bfseries{probability}   & Different & Mh (\textit{MhJK, MhCH})  & Mth (\textit{Mth}) \\
\end{tabular}
}
\end{table}

Based on the four basic CRC models, various estimators were developed. 
According to a recent systematic review \cite{liu2015adoption}, \textit{MhJK}, \textit{MhCH}, \textit{MtCH} are the three most frequently investigated and most effective estimators in software engineering.
Apart from that, we investigate another two estimators (i.e., \textit{M0} and \textit{Mth}) to ensure all four basic models are investigated.

\subsubsection{\textbf{How to Use in {\tool}}}

{\tool} treats each sample as a capture (or recapture).
Based on the \textit{bug arrival lookup table}, it then predicts the total number of bugs in the software using the CRC estimator.
This section first demonstrates how it works with \textit{Mth} estimator. 

\textit{Mth} estimator predicts the total number of bugs based on Equation \ref{equ_mth}, \ref{equ_mth3} \cite{Mthlee1996estimating}.
Table \ref{tab:Mth-estimator} shows the meaning of each variable, how to compute its value based on the bug arrival lookup table in Table \ref{tab:CRC-demo}.

\begin{equation}\label{equ_mth}
\footnotesize{ N = \frac{D}{C} + \frac{f_1}{C}\gamma^2 } \mbox{, } \footnotesize{ C = 1 - \frac{f_1}{\sum_{k=1}^{t}kf_k} } \\
\end{equation}

\begin{equation}\label{equ_mth3}
\footnotesize{ \gamma^2 = max\{{\frac{\frac{D}{C}\sum_{k}k(k-1)f_k}{2\sum\sum_{j<k}n_jn_k}-1, 0}\} }
\end{equation}

\begin{table}[!h]
\caption{Variables meaning and computation}
\label{tab:Mth-estimator}
\centering
\scriptsize
\begin{tabular}{p{0.35cm}|p{2.3cm}|p{3cm}|p{1cm}}
\hline
\textbf{Var.} & \textbf{Meaning} & \textbf{Computation based on bug arrival lookup table} & \textbf{Example value}\\
\hline
N & Predicted total number of bugs & & predicted value: \textbf{\textit{24}}\\
\hline
D & Actual number of bugs captured so far & Number of columns & 12 \\
\hline
t & Number of captures & Number of rows & 6 \\
\hline
$n_j$ & Number of bugs detected in each capture & Number of cells with \textit{1} in row \textit{j} & {3, 2, 2, 5, 6, 4}\\
\hline
$f_k$ & Number of bugs captured exactly $k$ times in all captures, i.e., $\sum f_i = D$  & Count the number of cells with \textit{1} in each column, and denote as $r_i$; $f_k$ is the number of $r_i$ with value \textit{k} & {1=7, 2=2, 3=2, 5=1}\\
\hline
\end{tabular}
\end{table}

For the usage of other four estimators, one can find the equation for estimating the total bugs from related work (i.e., \cite{M0laplace1783naissances} for \textit{M0}, \cite{MtCHchao1987estimating} for \textit{MtCH}, \cite{MhCHchao1988estimating} for \textit{MhCH}, and \cite{MhJKburnham1978estimation} for \textit{MhJK}).
The value assignments for the variables are the same as \textit{Mth}.
Due to space limit, we put the detailed illustration on our website.

\subsection{Predict Required Cost Using ARIMA}
\label{subsec:approach_cost_predict}

\subsubsection{\textbf{Background about ARIMA}}
\label{subsec:background_related_ARIMA}

ARIMA model is commonly used to model time series data to forecast the future values \cite{amin2013approach,chong1988analyzing,kemerer1999empirical,herraiz2007forecasting}.
It extends ARMA (Autoregressive Moving Average) model by allowing for non-stationary time series to be modeled, i.e., a time series whose statistical properties such as mean, variance, etc. are not constant 
over time.

A time series is said to be autoregressive moving average (ARMA) in nature 
with parameters $(p,q)$, if it takes the following form:
\begin{equation}
\label{eq:arima}
y_t=\sum_{i=1}^{p}\phi_i 
y_{t-i}+\sum_{i=1}^{q}\theta_i\epsilon_{t-i}+\epsilon_t
\end{equation}

Where $y_t$ is the current stationary observation, $y_{t-i}$ for \mbox{$i = 1, . . 
., p$}
are the past stationary observations, $\epsilon_t$ is the current error, and 
$\epsilon_{t-i}$
for $i = 1, . . ., q$ are the past errors. If this original time series 
$\{z_t\}$ is non-stationary, then $d$ differences can be done to transform it 
into a stationary one $\{y_t\}$. These differences can be viewed as a 
transformation denoted by $y_t = \triangledown^dz_t$, where 
$\triangledown^d=(1-B)^d$ where $B$ is known as a backshift operator. When 
this differencing operation is performed, it converts an ARMA (Autoregressive 
Moving Average) model into an ARIMA (Autoregressive Integrated Moving Average) 
model with parameters $(p,q,d)$.

\subsubsection{\textbf{How to Use in {\tool}}}

\begin{figure}[ht!]
\centering
\includegraphics[width=5cm]{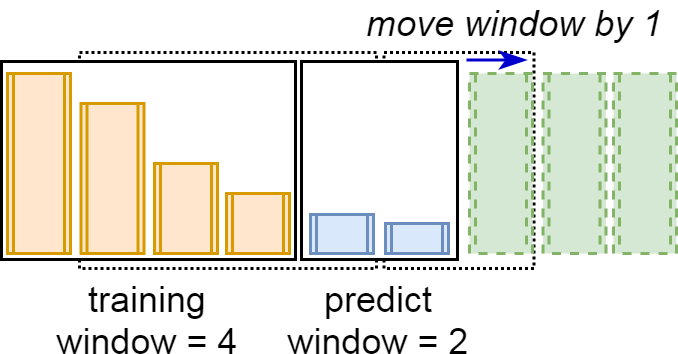}
\vspace{0.05in}
\caption{Illustrative example of ARIMA}
\label{fig:ARIMA}
\end{figure}

Figure \ref{fig:ARIMA} demonstrates how ARIMA is applied in predicting future trend of bug arrival. 
We treat the reports of each sample as a window, and obtain the number of unique bugs submitted in each sample from bug arrival lookup table. 
Then we use the former \textit{trainSize} windows to fit the ARIMA model and predict the number of bugs for the later \textit{predictSize} windows. 
When new window is formed with the newly-arrived reports, we move the window by 1 and obtain the newly predicted results. 

Suppose one want to know how much extra cost is required for achieving \textit{X\%} bugs. 
As we already know the predicted total number of bugs (Section \ref{subsec:approach_bug_predict}), we can figure out how many bugs should be detected in order to meet the test objective (i.e., \textit{X\%} bugs); suppose it is \textit{Y} bugs. 
Based on the prediction of ARIMA, we then obtain when the number of \textit{Y} bugs can be received, suppose it needs extra \textit{$K_i$} reports. 
In this way, we assume \textit{$K_i$} is the required cost for meeting the test objective.

\subsection{Apply {\tool} to Two Decision Scenarios in Crowdtesting}
\label{subsec:approach_application}

To demonstrate the usefulness of {\tool}, we generalize two typical decision scenarios in crowdtesting management, and illustrate its application to each scenario. 

\subsubsection{\textbf{Automating Task Closing Decision}}
\label{subsubsec:approach_application_auto}

The first scenario that can benefit from the prediction of total bugs of {\tool} (Section \ref{subsec:approach_bug_predict}) is decision automation of dynamic task closing.

As soon as a crowdtesting task begins, {\tool} can be applied to monitor the actual bug arrival, constantly update the bug arrival lookup table, as well as keep tracking of the percentage of bugs detected (i.e., the ratio of the number of submitted bugs so far over the predicted total bugs). 

In such scenario, different task close criteria can be customized in {\tool} so that it automatically closes the task when the specified criterion is met. 
For instance, a simple criterion would be to close the task when 100\% bugs have been detected in submitted reports.
Under this criterion, when {\tool} monitors 100\% bugs have received and the prediction remains unchanged for successive two captures, 
it determines the time, when the last report was received, as the close time; and would automatically close the crowdtesting task at run time. 
Note that the restriction of two successive captures is to ensure the stability of the prediction.
This is because our investigation reveals that without this restriction, {\tool} would occasionally fall into quite bad performance\footnote{We also experimented with other restrictions (i.e., 1 to 5), results turn out that restriction with 2 can obtain relative good and stable performance; hence we only present these results due to space limit.}.

{\tool} supports flexible customization of the close criteria. As an example, a task manager can set to close his/her tasks when 80\% bugs have been detected. Consequently, {\tool} will help to monitor and close the task by reacting to these customized close criteria.

\subsubsection{\textbf{Semi-Automation of Task Closing Trade-off Analysis}}

The second scenario that benefits from the prediction of required cost of {\tool} (Section \ref{subsec:approach_cost_predict}) is decision support of task closing trade-off analysis.

For example, suppose 90\% bugs have been reported at certain time, {\tool} can simultaneously reveal the estimated required cost for detecting an additional X\% bugs (i.e., 5\%), in order to achieve a higher bug detection level. Such cost-benefit related insights can provide managers with more confidence in making informed, actionable decision on whether to close immediately, if the required cost is too high to be worthwhile for additional X\% more detected bugs, or wait a little longer, if the required cost is acceptable and additional X\% detected bugs is desired.

\section{Experiment Design}
\label{sec:experiment}

\subsection{Research Questions}
\label{subsec:experiment_rq}

Four research questions are formulated to investigate the performance of the proposed {\tool}.

The first two research questions are centered around accuracy evaluation of the prediction of total bugs and required cost. 
Presumably, to support practical decision making, these underlying predictions should achieve high accuracy. 


\begin{itemize}
\item RQ1: To what degree can  {\tool} accurately predict total bugs?
\end{itemize}

\begin{itemize}
\item RQ2: To what degree can  {\tool} accurately predict required cost to achieve certain test objectives?
\end{itemize}

The next two research questions are focused on investigating the effectiveness of applying {\tool} in  the two typical scenarios (Section \ref{subsec:approach_application}), in which {\tool} is expected to alleviate current practices through automated and semi-automated decision support in managing crowdtesting tasks.

\begin{itemize}
\item RQ3: To what extent can {\tool} help to increase the effectiveness of crowdtesting through decision automation?
\end{itemize}

\begin{itemize}
\item RQ4: 
How {\tool} can be applied to facilitate the trade-off decisions about cost-effectiveness?
\end{itemize}

\subsection{Evaluation Metrics}
\label{subsec:experiment_metric}

We measure the \textbf{accuracy} of prediction based on \textbf{relative error}, which is the most commonly-used measure for accuracy \cite{wang2016towards,tan2015online,nam2017heterogeneous}.
It is applied in the prediction of total number of bugs (Section \ref{subsec:result_RQ1}) and required cost (Section \ref{subsec:result_RQ2}).


We measure the \textbf{cost-effectiveness} of close prediction (Section \ref{subsec:application_close}) based on two metrics, i.e. bug detection level (i.e. \textbf{\%bug}) and cost reduction (i.e. \textbf{\%reducedCost}).

\textbf{\%bug} is the percentage of bugs detected by the predicted close time. We treat the number of historical detected bugs as the total number.
The larger \textit{\%bug}, the better.

\textbf{\%reducedCost} is the percentage of saved cost by the predicted close time.
To derive this metric, we first obtain the percentage of reports submitted at the close time, in which we treat the number of historical submitted reports as the total number.
We suppose this is the percentage of consumed cost and \%reducedCost is derived using 1 minus the percentage of consumed cost.
The larger \textit{\%reducedCost}, the better.

Intuitively, an increase in \%bug would be accompanied with a decrease in \%reducedCost.
Motivated by the F1 (or F-Measure) in prediction approaches of software engineering \cite{nam2017heterogeneous,wang2017domain,wang2016towards}, we further derive an analogous metric \textbf{F1}, to measure the harmonic mean of \%bug and \%reducedCost as follows: 
\begin{equation}
\footnotesize{ F1 = \frac{2 \times \%bug \times \%reducedCost}{\%bug + \%reducedCost} }
\end{equation}

\subsection{Experimental Setup}
\label{subsec:experiment_setup}

For RQ1, we set up 19 checkpoints in the range of receiving 10\% to 100\% reports, with an increment interval of 5\% in between. 
At each checkpoint, we obtain the estimated total number of bugs at that time (see Section \ref{subsec:approach_bug_predict}).
Based on the ground truth of actual total bugs, we then figure out the relative error (Section \ref{subsec:experiment_metric}) in predicting the total bugs for each task.

For RQ2, we also set 19 checkpoints as RQ1. 
Different from RQ1, the checkpoints of RQ2 is based on the percentage of detected bugs, i.e. from 10\% bugs to 100\% bugs with an increment of 5\% in between.
At each checkpoint, we predict the required test cost (Section \ref{subsec:approach_cost_predict}) to achieve an additional 5\% bugs, i.e. target corresponding to the next checkpoint. 
For example, at the checkpoint when 80\% bugs have detected, we predict the required cost for achieving 85\% bugs. 
Based on the ground truth of actual required cost, we then figure out the relative error (Section \ref{subsec:experiment_metric}) in predicting required cost for each task.

For RQ3, we analyze the effectiveness of task closing automation with respect to five sample close criteria, i.e., close the task when 80\%, 85\%, 90\%, 95\%, or 100\% bugs have detected, respectively. These five close criteria are consistent with the commonly-used test completion criteria in software testing, and we believe the similar principles can be adopted in crowdtesting as well. 

For RQ4, we use several illustrative cases from experimental projects to show how {\tool} can help trade-off decisions. 


\subsection{Baselines}
\label{subsec:experiment_baselines}

To further evaluate the advantages of our proposed {\tool}, we compare it with two baselines. 

\textbf{Rayleigh}: This baseline is adopted from one of the most classical models for predicting the dynamic defect arrival in software measurement. Generally, it supposes the defect arrival data following the Rayleigh probability distribution \cite{kan2002metrics}. In this experiment, we implement code to fit specific Rayleigh curve (i.e. the derived Rayleigh model) based on each task's bug arrival data, and then predict the total bugs, as well as the future bug trend (and further obtain the required cost for certain test objective), using the derived Rayleigh model.

\textbf{Naive}: This baseline is designed to employ naive empirical results, i.e. the median obtained from the experimental dataset.  More specifically, for the prediction of total bugs, it uses the median total bugs from {\task} experimental tasks.
For required cost, it uses the median required cost from {\task} experimental tasks, in terms of the corresponding checkpoint (Section \ref{subsec:experiment_setup}).

\subsection{Parameter Tuning}
\label{subsec:experiment_parameter}

For each CRC estimator, the input parameter is \textit{smpSize}, which represents how many reports are considered in each capture.
To determine the value of this parameter, we random select 2/3 crowdtesting tasks to tune the parameter, and repeat the tuning for 1000 times to alleviate the randomness. 

In each tuning, for every candidate parameter value (we experiment from 2 to 30) and for each checkpoint, we obtain the median relative error for the prediction of total bugs (as shown in Table \ref{tab:testAdeStatis}) in terms of {\task} experimental tasks. 
Then for each candidate parameter value, we sum all absolute values of relative error for all checkpoints. 
We treat the parameter value, under which the smallest sum is obtained, as the best one. 
Finally, we use the parameter value which appears most frequently in the 1000 random experiments.
The tuned \textit{smpSize} values are respectively 8 for \textit{M0}, 8 for \textit{MtCH}, 6 for \textit{MhCH}, 3 for \textit{MhJK}, and 8 for \textit{Mth}.

For ARIMA model, we use the same method for deciding the best parameter value.
The tuned parameter values are as follows: \textit{smpSize} is 3, \textit{trainSize} is 10, \textit{p, q, and d} are respectively 5, 1, 0. 

\section{Results}
\label{sec:result}

\subsection{Answers to RQ1 : Accuracy of Total Bugs Prediction}
\label{subsec:result_RQ1}

\begin{figure}[t!]
\centering
\includegraphics[width=8cm]{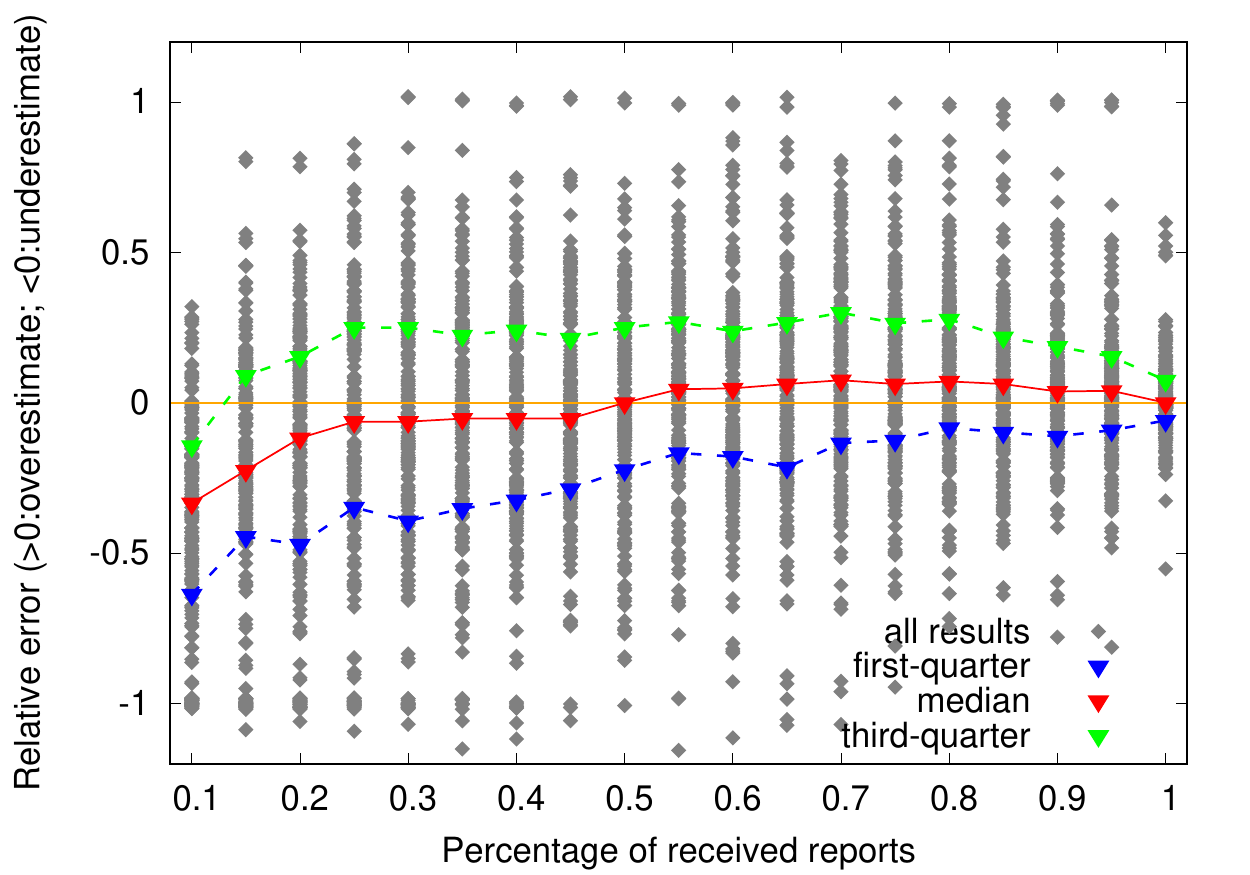}
\caption{Relative error of predicted total bugs of \textbf{MhJK} (RQ1). Note that we do not recommend this estimator since the (25-75)th range is not improving over most of the checkpoints.}
\label{fig:testAde-MhJK}
\vspace{-0.15in}
\end{figure}

\begin{figure}[t!]
\centering
\includegraphics[width=8cm]{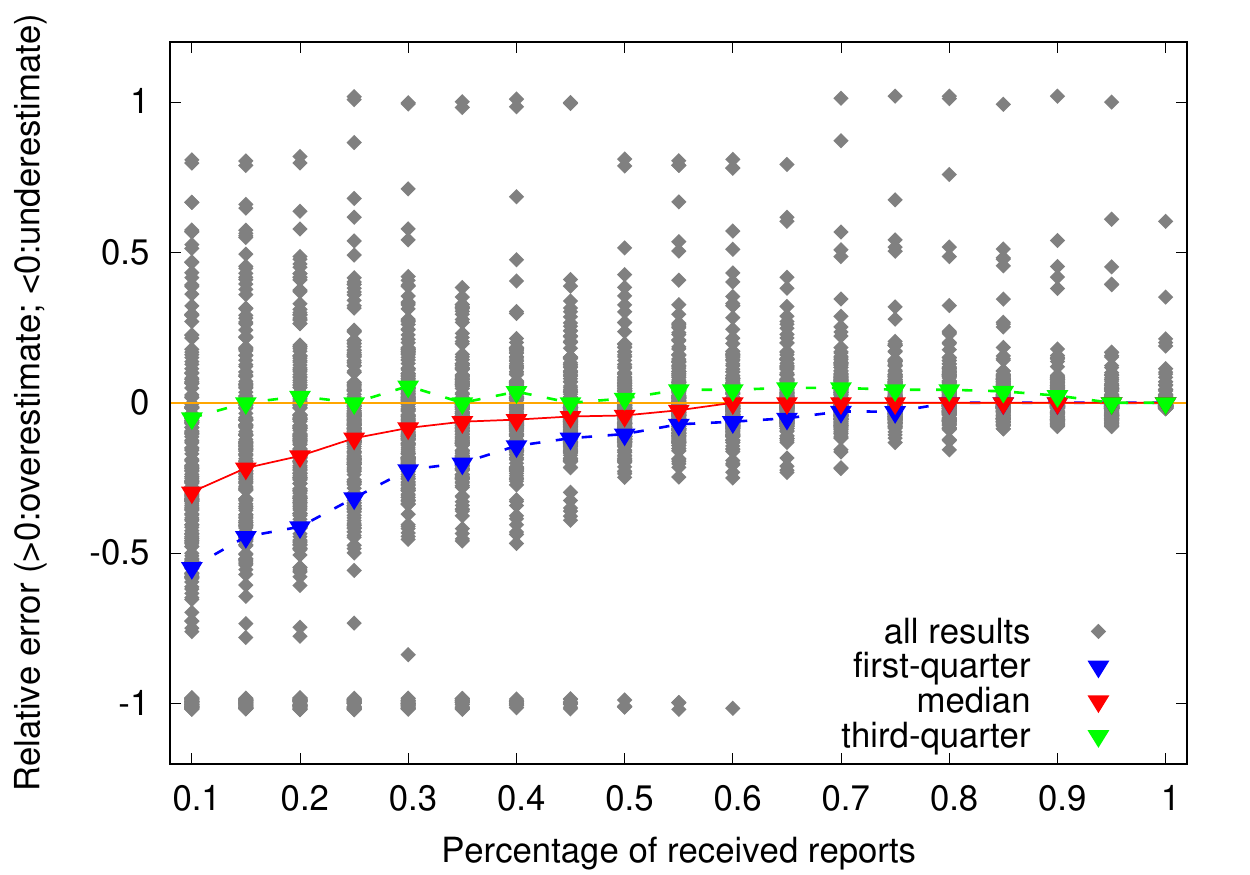}
\caption{Relative error of predicted total bugs of \textbf{Mth} (RQ1). We recommend this estimator since the (25-75)th range shrinks to very small values, very quickly, across the checkpoints.}
\label{fig:testAde-Mth}
\vspace{-0.15in}
\end{figure}



\begin{table*}[t!]
\scriptsize
\caption{Statistics for relative error of predicted total bugs (RQ1)}
\label{tab:testAdeStatis}
\centering
\scalebox{0.97}{
\begin{tabular}{p{0.6cm}|p{0.55cm}|p{0.55cm}|p{0.55cm}|p{0.55cm}|p{0.55cm}|p{0.55cm}|p{0.55cm}|p{0.55cm}|p{0.55cm}|p{0.55cm}|p{0.55cm}|p{0.55cm}|p{0.3cm}|p{0.3cm}|p{0.3cm}|p{0.3cm}|p{0.3cm}|p{0.3cm}|p{0.35cm}}
\hline
& \textbf{10\%} & \textbf{15\%} & \textbf{20\%} & \textbf{25\%} & \textbf{30\%} & \textbf{35\%} & \textbf{40\%} & \textbf{45\%} &  \textbf{50\%} & \textbf{55\%} & \textbf{60\%} & \textbf{65\%} & \textbf{70\%} & \textbf{75\%} & \textbf{80\%} & \textbf{85\%} & \textbf{90\%} & \textbf{95\%} & \textbf{100\%} \\
\hline
\hline
 & \multicolumn{19}{c}{\textbf{\textit{Median}}} \\ \hline 
M0 & -0.35  & -0.25  & -0.23  & -0.14  & -0.10  & -0.07  & \cellcolor{mypink}\textbf{\textit{-0.04}}  & \cellcolor{mypink}\textbf{\textit{-0.04}}  & -0.04  & -0.03  & -0.02  & \cellcolor{mypink}\textbf{\textit{0.00}}  & \cellcolor{mypink}\textbf{\textit{0.00}}  & \cellcolor{mypink}\textbf{\textit{0.00}}  & \cellcolor{mypink}\textbf{\textit{0.00}}  & \cellcolor{mypink}\textbf{\textit{0.00}}  & \cellcolor{mypink}\textbf{\textit{0.00}}  & \cellcolor{mypink}\textbf{\textit{0.00}}  & \cellcolor{mypink}\textbf{\textit{0.00}}  \\ \hline
MtCH & -0.40  & -0.32  & -0.28  & -0.18  & -0.15  & -0.10  & -0.07  & -0.06  & -0.05  & -0.03  & -0.02  & \cellcolor{mypink}\textbf{\textit{0.00}}  & \cellcolor{mypink}\textbf{\textit{0.00}}  & \cellcolor{mypink}\textbf{\textit{0.00}}  & \cellcolor{mypink}\textbf{\textit{0.00}}  & \cellcolor{mypink}\textbf{\textit{0.00}}  & \cellcolor{mypink}\textbf{\textit{0.00}}  & \cellcolor{mypink}\textbf{\textit{0.00}}  & \cellcolor{mypink}\textbf{\textit{0.00}}  \\ \hline
MhCH & -0.36  & -0.23  & \cellcolor{mypink}\textbf{\textit{-0.17}}  & -0.12  & -0.12  & -0.07  & \cellcolor{mypink}\textbf{\textit{-0.05}}  & -0.05  & \cellcolor{mypink}\textbf{\textit{-0.02}}  & \cellcolor{mypink}\textbf{\textit{0.00}}  & \cellcolor{mypink}\textbf{\textit{0.00}}  & \cellcolor{mypink}\textbf{\textit{0.00}}  & \cellcolor{mypink}\textbf{\textit{0.00}}  & \cellcolor{mypink}\textbf{\textit{0.00}}  & \cellcolor{mypink}\textbf{\textit{0.00}}  & \cellcolor{mypink}\textbf{\textit{0.00}}  & \cellcolor{mypink}\textbf{\textit{0.00}}  & \cellcolor{mypink}\textbf{\textit{0.00}}  & \cellcolor{mypink}\textbf{\textit{0.00}}  \\ \hline
MhJK & \cellcolor{mypink}\textbf{\textit{-0.33}}  & \cellcolor{mypink}\textbf{\textit{-0.22}}  & \cellcolor{mypink}\textbf{\textit{-0.11}}  & \cellcolor{mypink}\textbf{\textit{-0.06}}  & \cellcolor{mypink}\textbf{\textit{-0.06}}  & \cellcolor{mypink}\textbf{\textit{-0.05}}  & -0.06  & -0.05  & \cellcolor{mypink}\textbf{\textit{0.00}}  & 0.04  & 0.04  & 0.06  & 0.08  & 0.06  & 0.06  & 0.06  & 0.03  & 0.04  & \cellcolor{mypink}\textbf{\textit{0.00}}  \\ \hline
Mth & \cellcolor{mypink}\textbf{\textit{-0.29}}  & \cellcolor{mypink}\textbf{\textit{-0.21}}  & \cellcolor{mypink}\textbf{\textit{-0.17}}  & \cellcolor{mypink}\textbf{\textit{-0.11}}  & \cellcolor{mypink}\textbf{\textit{-0.08}}  & \cellcolor{mypink}\textbf{\textit{-0.06}}  & \cellcolor{mypink}\textbf{\textit{-0.05}}  & \cellcolor{mypink}\textbf{\textit{-0.04}}  & -0.04  & \cellcolor{mypink}\textbf{\textit{-0.02}}  & \cellcolor{mypink}\textbf{\textit{0.00}}  & \cellcolor{mypink}\textbf{\textit{0.00}}  & \cellcolor{mypink}\textbf{\textit{0.00}}  & \cellcolor{mypink}\textbf{\textit{0.00}}  & \cellcolor{mypink}\textbf{\textit{0.00}}  & \cellcolor{mypink}\textbf{\textit{0.00}}  & \cellcolor{mypink}\textbf{\textit{0.00}}  & \cellcolor{mypink}\textbf{\textit{0.00}}  & \cellcolor{mypink}\textbf{\textit{0.00}}  \\ \hline
 & \multicolumn{19}{c}{\textbf{\textit{Standard deviation}}} \\ \hline 
M0 & 0.40  & \cellcolor{mypink}\textbf{\textit{0.40}}  & 0.40  & 0.40  & 0.38  & 0.37  & 0.34  & 0.32  & 0.29  & 0.26  & 0.23  & 0.21  & 0.20  & 0.17  & \cellcolor{mypink}\textbf{\textit{0.13}}  & 0.12  & \cellcolor{mypink}\textbf{\textit{0.07}}  & \cellcolor{mypink}\textbf{\textit{0.05}}  & \cellcolor{mypink}\textbf{\textit{0.03}}  \\ \hline
MtCH & \cellcolor{mypink}\textbf{\textit{0.31}}  & \cellcolor{mypink}\textbf{\textit{0.32}}  & \cellcolor{mypink}\textbf{\textit{0.32}}  & \cellcolor{mypink}\textbf{\textit{0.32}}  & \cellcolor{mypink}\textbf{\textit{0.29}}  & \cellcolor{mypink}\textbf{\textit{0.26}}  & \cellcolor{mypink}\textbf{\textit{0.21}}  & \cellcolor{mypink}\textbf{\textit{0.17}}  & \cellcolor{mypink}\textbf{\textit{0.13}}  & \cellcolor{mypink}\textbf{\textit{0.14}}  & \cellcolor{mypink}\textbf{\textit{0.13}}  & \cellcolor{mypink}\textbf{\textit{0.09}}  & \cellcolor{mypink}\textbf{\textit{0.10}}  & \cellcolor{mypink}\textbf{\textit{0.08}}  & \cellcolor{mypink}\textbf{\textit{0.09}}  & \cellcolor{mypink}\textbf{\textit{0.08}}  & \cellcolor{mypink}\textbf{\textit{0.07}}  & \cellcolor{mypink}\textbf{\textit{0.07}}  & \cellcolor{mypink}\textbf{\textit{0.04}}  \\ \hline
MhCH & \cellcolor{mypink}\textbf{\textit{0.36}}  & 0.40  & \cellcolor{mypink}\textbf{\textit{0.39}}  & 0.39  & 0.38  & \cellcolor{mypink}\textbf{\textit{0.31}}  & 0.28  & 0.25  & 0.20  & 0.19  & 0.22  & 0.20  & 0.21  & 0.16  & 0.20  & 0.18  & 0.18  & 0.17  & 0.09  \\ \hline
MhJK & 0.37  & 0.42  & 0.44  & 0.45  & 0.44  & 0.41  & 0.39  & 0.36  & 0.35  & 0.34  & 0.36  & 0.35  & 0.33  & 0.31  & 0.29  & 0.28  & 0.27  & 0.25  & 0.14  \\ \hline
Mth & 0.42  & 0.42  & 0.40  & \cellcolor{mypink}\textbf{\textit{0.38}}  & \cellcolor{mypink}\textbf{\textit{0.35}}  & 0.33  & \cellcolor{mypink}\textbf{\textit{0.27}}  & \cellcolor{mypink}\textbf{\textit{0.24}}  & \cellcolor{mypink}\textbf{\textit{0.19}}  & \cellcolor{mypink}\textbf{\textit{0.19}}  & \cellcolor{mypink}\textbf{\textit{0.16}}  & \cellcolor{mypink}\textbf{\textit{0.11}}  & \cellcolor{mypink}\textbf{\textit{0.13}}  & \cellcolor{mypink}\textbf{\textit{0.12}}  & 
\cellcolor{mypink}\textbf{\textit{0.13}} & \cellcolor{mypink}\textbf{\textit{0.11}}  & 0.10  & 
\cellcolor{mypink}\textbf{\textit{0.07}} & \cellcolor{mypink}\textbf{\textit{0.04}}  \\ \hline

\end{tabular}
}
\end{table*}

\begin{table*}[t!]
\scriptsize
\caption{Comparison with baselines in relative error of predicted total bugs (RQ1)}
\label{tab:testAdeStatis_baseline}
\centering
\scalebox{0.9}{
\begin{tabular}{p{0.8cm}|p{0.55cm}|p{0.55cm}|p{0.55cm}|p{0.55cm}|p{0.55cm}|p{0.55cm}|p{0.55cm}|p{0.55cm}|p{0.55cm}|p{0.55cm}|p{0.55cm}|p{0.55cm}|p{0.55cm}|p{0.55cm}|p{0.55cm}|p{0.55cm}|p{0.55cm}|p{0.55cm}|p{0.55cm}}
\hline
& \textbf{10\%} & \textbf{15\%} & \textbf{20\%} & \textbf{25\%} & \textbf{30\%} & \textbf{35\%} & \textbf{40\%} & \textbf{45\%} &  \textbf{50\%} & \textbf{55\%} & \textbf{60\%} & \textbf{65\%} & \textbf{70\%} & \textbf{75\%} & \textbf{80\%} & \textbf{85\%} & \textbf{90\%} & \textbf{95\%} & \textbf{100\%} \\
\hline
\hline
 & \multicolumn{19}{c}{\textbf{\textit{Median}}} \\ \hline 
{\tool} & -0.29  & \cellcolor{mypink}\textbf{\textit{-0.21}}  & \cellcolor{mypink}\textbf{\textit{-0.17}}  & \cellcolor{mypink}\textbf{\textit{-0.11}}  & \cellcolor{mypink}\textbf{\textit{-0.08}}  & \cellcolor{mypink}\textbf{\textit{-0.06}}  & \cellcolor{mypink}\textbf{\textit{-0.05}}  & \cellcolor{mypink}\textbf{\textit{-0.04}}  & \cellcolor{mypink}\textbf{\textit{-0.04}}  & \cellcolor{mypink}\textbf{\textit{-0.02}}  & \cellcolor{mypink}\textbf{\textit{0.00}}  & \cellcolor{mypink}\textbf{\textit{0.00}}  & \cellcolor{mypink}\textbf{\textit{0.00}}  & \cellcolor{mypink}\textbf{\textit{0.00}}  & \cellcolor{mypink}\textbf{\textit{0.00}}  & \cellcolor{mypink}\textbf{\textit{0.00}}  & \cellcolor{mypink}\textbf{\textit{0.00}}  & \cellcolor{mypink}\textbf{\textit{0.00}}  & \cellcolor{mypink}\textbf{\textit{0.00}}  \\ \hline
Rayleigh & -0.50  & -0.43  & -0.35  & -0.24  & -0.23  & -0.22  & -0.18  & -0.15  & -0.14  & -0.13  & -0.13  & -0.13  & -0.13  & -0.12  & -0.12  & -0.07  & \cellcolor{mypink}\textbf{\textit{0.00}}  & \cellcolor{mypink}\textbf{\textit{0.00}}  & \cellcolor{mypink}\textbf{\textit{0.00}}  \\ \hline
Naive & \cellcolor{mypink}\textbf{\textit{-0.26}}  & -0.26  & -0.26  & -0.26  & -0.26  & -0.26  & -0.26  & -0.26  & -0.26  & -0.26  & -0.26  & -0.26  & -0.26  & -0.26  & -0.26  & -0.26  & -0.26  & -0.26  &-0.26  \\ \hline
 & \multicolumn{19}{c}{\textbf{\textit{Standard deviation}}} \\ \hline 
{\tool} & 0.42  & 0.42  & 0.40  & 0.38  & 0.35  & 0.33  & \cellcolor{mypink}\textbf{\textit{0.27}}  & \cellcolor{mypink}\textbf{\textit{0.24}}  & \cellcolor{mypink}\textbf{\textit{0.19}}  & \cellcolor{mypink}\textbf{\textit{0.19}}  & \cellcolor{mypink}\textbf{\textit{0.16}}  & \cellcolor{mypink}\textbf{\textit{0.11}}  & \cellcolor{mypink}\textbf{\textit{0.13}}  & \cellcolor{mypink}\textbf{\textit{0.12}}  & \cellcolor{mypink}\textbf{\textit{0.13}}  & \cellcolor{mypink}\textbf{\textit{0.11}}  & \cellcolor{mypink}\textbf{\textit{0.10}}  & \cellcolor{mypink}\textbf{\textit{0.10}}  & \cellcolor{mypink}\textbf{\textit{0.06}}  \\ \hline
Rayleigh & \cellcolor{mypink}\textbf{\textit{0.23}}  & \cellcolor{mypink}\textbf{\textit{0.26}}  & \cellcolor{mypink}\textbf{\textit{0.30}}  & \cellcolor{mypink}\textbf{\textit{0.30}}  & \cellcolor{mypink}\textbf{\textit{0.29}}  & \cellcolor{mypink}\textbf{\textit{0.30}}  & 0.35  & 0.38  & 0.40  & 0.52  & 0.52  & 0.63  & 0.66  & 0.71  & 0.70  & 0.64  & 0.45  & 0.35  & 0.27  \\ \hline
Naive & 0.48  & 0.48  & 0.48  & 0.48  & 0.48  & 0.48  & 0.48  & 0.48  & 0.48  & 0.48  & 0.48  & 0.48  & 0.48  & 0.48  & 0.48  & 0.48  & 0.48  & 0.48  & 0.48  \\ \hline

\end{tabular}
}
\end{table*}

\begin{table*}[t!]
\scriptsize
\caption{Results of Mann-Whitney U Test for relative error of predicted total bugs (RQ1)}
\label{tab:testAdeStatis-baseline-test}
\centering\begin{tabular}{p{2.4cm}|p{0.35cm}|p{0.35cm}|p{0.35cm}|p{0.35cm}|p{0.35cm}|p{0.35cm}|p{0.35cm}|p{0.35cm}|p{0.35cm}|p{0.35cm}|p{0.35cm}|p{0.35cm}|p{0.35cm}|p{0.35cm}|p{0.35cm}|p{0.35cm}|p{0.35cm}|p{0.35cm}|p{0.45cm}}
\hline
& \textbf{10\%} & \textbf{15\%} & \textbf{20\%} & \textbf{25\%} & \textbf{30\%} & \textbf{35\%} & \textbf{40\%} & \textbf{45\%} &  \textbf{50\%} & \textbf{55\%} & \textbf{60\%} & \textbf{65\%} & \textbf{70\%} & \textbf{75\%} & \textbf{80\%} & \textbf{85\%} & \textbf{90\%} & \textbf{95\%} & \textbf{100\%} \\ 
\hline
{\tool} vs. Rayleigh & 0.00\tiny{*} & 0.00\tiny{*} & 0.00\tiny{*} & 0.00\tiny{*} & 0.00\tiny{*} & 0.00\tiny{*} & 0.00\tiny{*} & 0.00\tiny{*} & 0.00\tiny{*} & 0.00\tiny{*} & 0.00\tiny{*} & 0.00\tiny{*} & 0.00\tiny{*} & 0.00\tiny{*} & 0.00\tiny{*} & 0.00\tiny{*} & 0.00\tiny{*} & 0.00\tiny{*} & 0.00\tiny{*} \\ \hline
{\tool} vs. Naive & 0.02\tiny{*} & 0.81 & 0.43 & 0.01\tiny{*} & 0.00\tiny{*} & 0.00\tiny{*} & 0.00\tiny{*} & 0.00\tiny{*} & 0.00\tiny{*} & 0.00\tiny{*} & 0.00\tiny{*} & 0.00\tiny{*} & 0.00\tiny{*} & 0.00\tiny{*} & 0.00\tiny{*} & 0.00\tiny{*} & 0.00\tiny{*} & 0.00\tiny{*} & 0.00\tiny{*} \\ \hline
Rayleigh vs. Naive & 0.00\tiny{*} & 0.00\tiny{*} & 0.00\tiny{*} & 0.46 & 0.97 & 0.53 & 0.10 & 0.00\tiny{*} & 0.00\tiny{*} & 0.00\tiny{*} & 0.00\tiny{*} & 0.00\tiny{*} & 0.00\tiny{*} & 0.00\tiny{*} & 0.00\tiny{*} & 0.00\tiny{*} & 0.00\tiny{*} & 0.00\tiny{*} & 0.00\tiny{*} \\ \hline
\end{tabular}
\end{table*}

Table \ref{tab:testAdeStatis} demonstrates the median and standard deviation for the \textit{relative error} of predicted total bugs for all five CRC estimators, corresponding to all checkpoints.
We highlight (in italic font and red color) methods which have the best two performance with respect to each checkpoint. 
Due to space limit, we only present the detailed performance for \textit{MhJK} (the worst estimator) and \textit{Mth} (the best estimator) in Figure \ref{fig:testAde-MhJK}, \ref{fig:testAde-Mth}.

From Table \ref{tab:testAdeStatis} and Figure \ref{fig:testAde-Mth}, we can see that, the predicted total number of bugs becomes more close to the actual total number of bugs (i.e., the \textit{relative error} decreases) towards the end of the tasks. 
Among the five estimators, \textit{Mth} and \textit{MhCH} have the smallest median \textit{relative error} for most checkpoints.
But the variance of \textit{MhCH} is much larger than that of \textit{Mth}.
Hence, estimator \textit{Mth} is more preferred because of its relatively higher stability and accurate prediction in total number of bugs. 
In the following experiments, if not specially mentioned, the results are referring to those generated from {\tool} with \textit{Mth} estimator due to space limit.


\textbf{Comparison With Baselines:} 
Table \ref{tab:testAdeStatis_baseline} compares the prediction accuracy of {\tool} and the two baselines, in terms of the median and standard deviation of \textit{relative error}. The columns correspond to different checkpoints, and the best performer under each checkpoint are highlighted. Table \ref{tab:testAdeStatis-baseline-test} summarizes the results of Mann-Whitney U Test for the \textit{relative error} of predicted total bugs between each two methods. It shows that the {\tool} significantly outperforms the two baselines (with p-value \textless 0.05), especially during the later stage (i.e. after the 40\% checkpoint) of the crowdtesting tasks. 

\textbf{Answers to RQ1:} 
{\tool} with the best estimator \textit{Mth} is surprisingly accurate in predicting the total bugs in crowdtesting, and significantly outperforms the two baselines. More specifically, the median of predicted total bugs is equal with the ground truth (i.e., median \textit{relative error} is 0). 
Better yet, the standard deviation is about 10\% to 20\% during the latter half of the process.

\subsection{Answers to RQ2 : Accuracy of Required Cost Prediction}
\label{subsec:result_RQ2}


\begin{table*}[t!]
\scriptsize
\caption{Statistics for relative error of predicted required cost (RQ2)}
\label{tab:future_cost}
\centering\begin{tabular}{p{1.5cm}|p{0.40cm}|p{0.40cm}|p{0.40cm}|p{0.40cm}|p{0.40cm}|p{0.40cm}|p{0.40cm}|p{0.40cm}|p{0.40cm}|p{0.40cm}|p{0.40cm}|p{0.40cm}|p{0.40cm}|p{0.40cm}|p{0.40cm}|p{0.40cm}|p{0.40cm}|p{0.40cm}|p{0.55cm}} \hline 
& \textbf{10\%} & \textbf{15\%} & \textbf{20\%} & \textbf{25\%} & \textbf{30\%} & \textbf{35\%} & \textbf{40\%} & \textbf{45\%} &  \textbf{50\%} & \textbf{55\%} & \textbf{60\%} & \textbf{65\%} & \textbf{70\%} & \textbf{75\%} & \textbf{80\%} & \textbf{85\%} & \textbf{90\%} & \textbf{95\%} & \textbf{100\%} \\
 \hline 
 \hline 
 & \multicolumn{19}{c}{\textbf{\textit{Median}}} \\ \hline 
{\tool} & \cellcolor{mypink}\textbf{\textit{0.33}}  & \cellcolor{mypink}\textbf{\textit{0.20}}  & \cellcolor{mypink}\textbf{\textit{0.13}}  & \cellcolor{mypink}\textbf{\textit{0.12}}  & \cellcolor{mypink}\textbf{\textit{0.11}}  & \cellcolor{mypink}\textbf{\textit{0.05}}  & \cellcolor{mypink}\textbf{\textit{0.05}}  & \cellcolor{mypink}\textbf{\textit{0.04}}  & \cellcolor{mypink}\textbf{\textit{0.04}}  & \cellcolor{mypink}\textbf{\textit{0.03}}  & \cellcolor{mypink}\textbf{\textit{0.03}}  & \cellcolor{mypink}\textbf{\textit{0.02}}  & \cellcolor{mypink}\textbf{\textit{0.02}}  & \cellcolor{mypink}\textbf{\textit{0.02}}  & \cellcolor{mypink}\textbf{\textit{0.01}}  & \cellcolor{mypink}\textbf{\textit{0.00}}  & \cellcolor{mypink}\textbf{\textit{0.01}}  & \cellcolor{mypink}\textbf{\textit{0.00}}  & -0.02  \\ \hline
Rayleigh & 1.01  & 0.67  & 0.42  & 0.33  & 0.25  & 0.15  & 0.13  & 0.10  & 0.11  & 0.23  & 0.17  & 0.12  & 0.19  & 0.20  & 0.18  & 0.10  & 0.09  & 0.13  & \cellcolor{mypink}\textbf{\textit{0.01}}  \\ \hline
Naive & \cellcolor{mypink}\textbf{\textit{0.33}}  & \cellcolor{mypink}\textbf{\textit{0.20}}  & \cellcolor{mypink}\textbf{\textit{0.13}}  & \cellcolor{mypink}\textbf{\textit{0.12}}  & \cellcolor{mypink}\textbf{\textit{0.11}}  & \cellcolor{mypink}\textbf{\textit{0.05}}  & \cellcolor{mypink}\textbf{\textit{0.05}}  & \cellcolor{mypink}\textbf{\textit{0.04}}  & \cellcolor{mypink}\textbf{\textit{0.04}}  & 0.05  & 0.06  & 0.05  & 0.03  & 0.05  & 0.06  & 0.05  & 0.04  & 0.07  & 0.07  \\ \hline
 & \multicolumn{19}{c}{\textbf{\textit{Standard deviation}}} \\ \hline 
{\tool} & \cellcolor{mypink}\textbf{\textit{0.81}}  & \cellcolor{mypink}\textbf{\textit{0.61}}  & \cellcolor{mypink}\textbf{\textit{0.38}}  & \cellcolor{mypink}\textbf{\textit{0.49}}  & \cellcolor{mypink}\textbf{\textit{0.28}}  & \cellcolor{mypink}\textbf{\textit{0.27}}  & \cellcolor{mypink}\textbf{\textit{0.22}}  & \cellcolor{mypink}\textbf{\textit{0.23}}  & \cellcolor{mypink}\textbf{\textit{0.15}}  & \cellcolor{mypink}\textbf{\textit{0.15}}  & \cellcolor{mypink}\textbf{\textit{0.16}}  & \cellcolor{mypink}\textbf{\textit{0.14}}  & \cellcolor{mypink}\textbf{\textit{0.13}}  & \cellcolor{mypink}\textbf{\textit{0.16}}  & \cellcolor{mypink}\textbf{\textit{0.17}}  & \cellcolor{mypink}\textbf{\textit{0.12}}  & \cellcolor{mypink}\textbf{\textit{0.15}}  & \cellcolor{mypink}\textbf{\textit{0.23}}  & \cellcolor{mypink}\textbf{\textit{0.16}}  \\ \hline
Rayleigh & 1.33  & 0.94  & 0.63  & 0.71  & 0.40  & 0.37  & 0.29  & 0.30  & 0.18  & 1.05  & 1.01  & 0.86  & 0.72  & 0.70  & 0.59  & 0.53  & 0.46  & 0.46  & 0.27  \\ \hline
Naive & \cellcolor{mypink}\textbf{\textit{0.81}}  & \cellcolor{mypink}\textbf{\textit{0.61}}  & \cellcolor{mypink}\textbf{\textit{0.38}}  & \cellcolor{mypink}\textbf{\textit{0.49}}  & \cellcolor{mypink}\textbf{\textit{0.28}}  & \cellcolor{mypink}\textbf{\textit{0.27}}  & \cellcolor{mypink}\textbf{\textit{0.22}}  & \cellcolor{mypink}\textbf{\textit{0.23}}  & \cellcolor{mypink}\textbf{\textit{0.15}}  & 0.16  & 0.17  & 0.15  & 0.16  & 0.17  & 0.19  & \cellcolor{mypink}\textbf{\textit{0.12}}  & 0.16 & 0.24  & 0.18  \\ \hline
\end{tabular}
\end{table*}

\begin{table*}[t!]
\scriptsize
\caption{Results of Mann-Whitney U Test for relative error of predicted required cost (RQ2)}
\label{tab:future_cost_test}
\centering\begin{tabular}{p{2.4cm}|p{0.35cm}|p{0.35cm}|p{0.35cm}|p{0.35cm}|p{0.35cm}|p{0.35cm}|p{0.35cm}|p{0.35cm}|p{0.35cm}|p{0.35cm}|p{0.35cm}|p{0.35cm}|p{0.35cm}|p{0.35cm}|p{0.35cm}|p{0.35cm}|p{0.35cm}|p{0.35cm}|p{0.45cm}}\hline
& \textbf{10\%} & \textbf{15\%} & \textbf{20\%} & \textbf{25\%} & \textbf{30\%} & \textbf{35\%} & \textbf{40\%} & \textbf{45\%} &  \textbf{50\%} & \textbf{55\%} & \textbf{60\%} & \textbf{65\%} & \textbf{70\%} & \textbf{75\%} & \textbf{80\%} & \textbf{85\%} & \textbf{90\%} & \textbf{95\%} & \textbf{100\%} \\
\hline 
{\tool} vs. Rayleigh & 0.00\tiny{*} & 0.00\tiny{*} & 0.00\tiny{*} & 0.00\tiny{*} & 0.00\tiny{*} & 0.00\tiny{*} & 0.00\tiny{*} & 0.00\tiny{*} & 0.00\tiny{*} & 0.00\tiny{*} & 0.00\tiny{*} & 0.00\tiny{*} & 0.00\tiny{*} & 0.00\tiny{*} & 0.00\tiny{*} & 0.00\tiny{*} & 0.00\tiny{*} & 0.00\tiny{*} & 0.00\tiny{*}\\ \hline
{\tool} vs. Naive & 1.00 & 0.97 & 1.00 & 1.00 & 1.00 & 1.00 & 0.94 & 1.00 & 0.96 & 0.01\tiny{*} & 0.00\tiny{*} & 0.00\tiny{*} & 0.11 & 0.01\tiny{*} & 0.00\tiny{*} & 0.00\tiny{*} & 0.00\tiny{*} & 0.00\tiny{*} & 0.00\tiny{*}\\ \hline
Rayleigh vs. Naive & 0.00\tiny{*} & 0.00\tiny{*} & 0.00\tiny{*} & 0.00\tiny{*} & 0.00\tiny{*} & 0.00\tiny{*} & 0.00\tiny{*} & 0.00\tiny{*} & 0.00\tiny{*} & 0.00\tiny{*} & 0.00\tiny{*} & 0.00\tiny{*} & 0.00\tiny{*} & 0.00\tiny{*} & 0.00\tiny{*} & 0.00\tiny{*} & 0.00\tiny{*} & 0.00\tiny{*} & 0.12\\ \hline
\end{tabular}
\end{table*}

Table \ref{tab:future_cost} summarizes the comparison of median and standard deviation of the \textit{relative error} of predicted required cost across {\tool} and the two baselines, with columns corresponding to different checkpoints.
We highlight the methods with the best performance under each checkpoint. 
Table \ref{tab:future_cost_test} additionally presents the results from the Mann-Whitney U Test between each pair.

As indicated by the decreasing median \textit{relative error} in Table \ref{tab:future_cost}, the prediction of required cost becomes increasingly accurate for later checkpoints.
For example, after 50\% checkpoint, the median \textit{relative error} of predicted cost is lower than 3\%, with about 15\% standard deviation. 
This implies that {\tool} can effectively predict the required cost to targeted test objectives.


\textbf{Comparison With Baselines:} 
We can see that the median and standard deviation of \textit{relative error} for two baselines are worse than \textit{\tool} during the second half of the task process. Observed from Table \ref{tab:future_cost_test}, the difference between the proposed {\tool} and two baselines is significant during the second half of crowdtesting process (p-value \textless 0.05). This further signifies the advantages of the proposed {\tool}.

\textbf{Answers to RQ2:} {\tool} can predict the required test cost within averagely 3\% \textit{relative error} for later stage of crowdtesting (i.e. after 50\% checkpoint).

\subsection{Answers to RQ3: Effectiveness of Task Closing Automation }
\label{subsec:application_close}

\begin{figure}[t!]
\centering
\includegraphics[width=8.5cm]{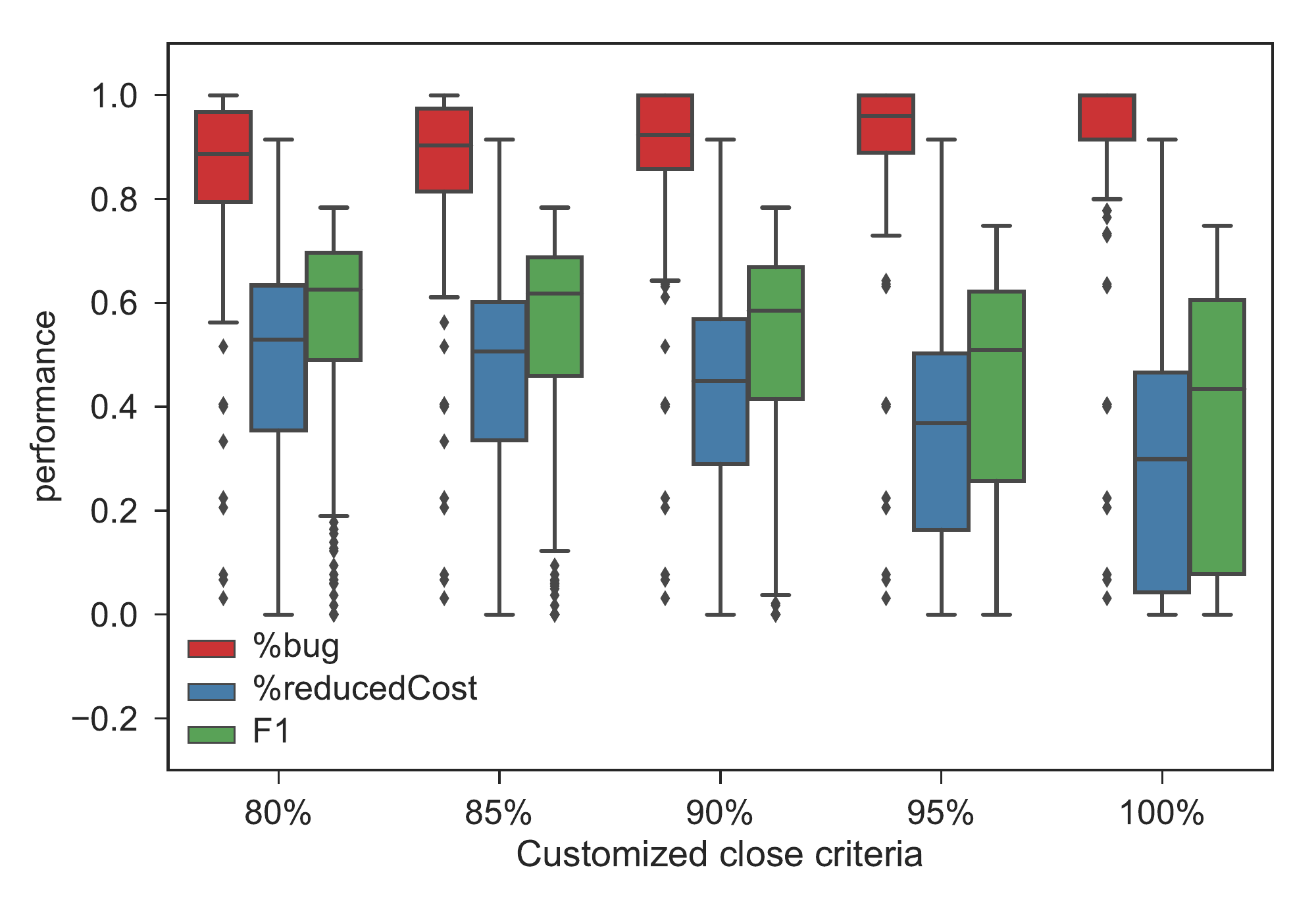}
\caption{Task closing automation performance (RQ3)}
\label{fig:costEff-combine}
\end{figure}

Figure \ref{fig:costEff-combine} shows the distribution of \textit{\%bug}, \textit{\%reducedCost}, and \textit{F1} for five customized close criteria for {\task} experimental tasks.
Table \ref{tab:costEffStatis-combine} lists their median and standard deviation. 

Let us first look at the last row in Table \ref{tab:costEffStatis-combine}, which reflects a close criterion of 100\% bugs being detected (i.e., most commonly-used setup).
The results indicate that a median of 100\% bugs can be detected with 29.9\% median cost reduction. This suggests an additional 30\% more cost-effectiveness for managers if equiped with such a decision automation tool as {\tool} to monitor and close tasks automatically at run-time. 
The reduced cost is a tremendous figure when considering the large number of tasks delivered in a crowdtesting platform. In addition, the standard deviation is relatively low, further signifying the stability of {\tool} in close automation.

We then shift our focus on other four customized close criteria (i.e., 80\%, 85\%, 90\%, and 95\% in terms of percentage of detected bugs).
We can observe that for each close criterion, the median \textit{\%bug} generated from {\tool} is very close to the targeted close criterion, with small standard deviation. 
Among these close criteria, 36\% to 52\% cost can be saved, which further signify the effectiveness of {\tool}.

We also notice that, the median \textit{\%bug} is a little larger than the customized close criterion. 
For example, if the project manager hopes to close the task when 90\% bugs detected, a median of 92\% bugs have submitted at the predicted close time.
This implies, in most cases, the close prediction produced by {\tool} do not have the risk of insufficient testing. 
Furthermore, we have talked with the project managers and they thought, detecting slightly more bugs (even with less reduced cost) is always better than detecting fewer bugs (with more reduced cost).
This is because \textit{\%bug} is more like the constraint, while \textit{\%reducedCost} is only the bonus.

We also analyze the reason for this phenomenon. 
It is mainly because, before suggesting close, our approach requires the predicted total bugs remain unchanged for two successive captures (Section \ref{subsubsec:approach_application_auto}). 
This restriction is to alleviate the risk of insufficient percentage of detected bugs. 
Besides, this is also because we treat a sample of reports as the unit during the prediction, which can also potentially result in the close time being a little later than the customized close time.

\begin{table}[t!]
\scriptsize
\caption{Statistics of task closing automation (RQ3)}
\label{tab:costEffStatis-combine}
\centering
\begin{tabular}{p{1.2cm}|p{0.7cm}p{0.7cm}p{0.7cm}|p{0.7cm}p{0.7cm}p{0.7cm}}
\hline
& \multicolumn{3}{c|}{Median} & \multicolumn{3}{c}{Standard deviation} \\
\hline
 &  \%B & \%R & F1 & \%B & \%R & F1  \\
\hline
80\% & 0.886  & 0.529  & 0.625  & 0.172  & 0.216  & 0.210  \\ \hline
85\% & 0.903  & 0.506  & 0.618  & 0.170  & 0.216  & 0.219  \\ \hline
90\% & 0.923  & 0.449  & 0.584  & 0.163  & 0.217  & 0.233  \\ \hline
95\% & 0.960  & 0.368  & 0.508  & 0.163  & 0.224  & 0.250  \\ \hline
100\% & 1.000  & 0.299  & 0.433  & 0.173  & 0.235  & 0.268  \\ \hline

\end{tabular}
\end{table}

\textbf{Answers to RQ3:} The automation of task closing by {\tool} can make crowdtesting more cost-effective, i.e., a median of 100\% bugs can be detected with 30\% saved cost.

\subsection{Answers to RQ4: Trade-off Decision Support}
\label{subsec:application_tradeoff}

Considering the large number of tasks under management at the same time, a typical trade-off scenario is to strategically allocate limited testing budgets among the tasks.  To reflect such trade-off context, we randomly  pick a time and slice the experimental dataset to retrieve all tasks under testing at that time, then examine the cost-effectiveness of more testing on those tasks.

\begin{figure}[t!]
\centering
\includegraphics[width=8.5cm]{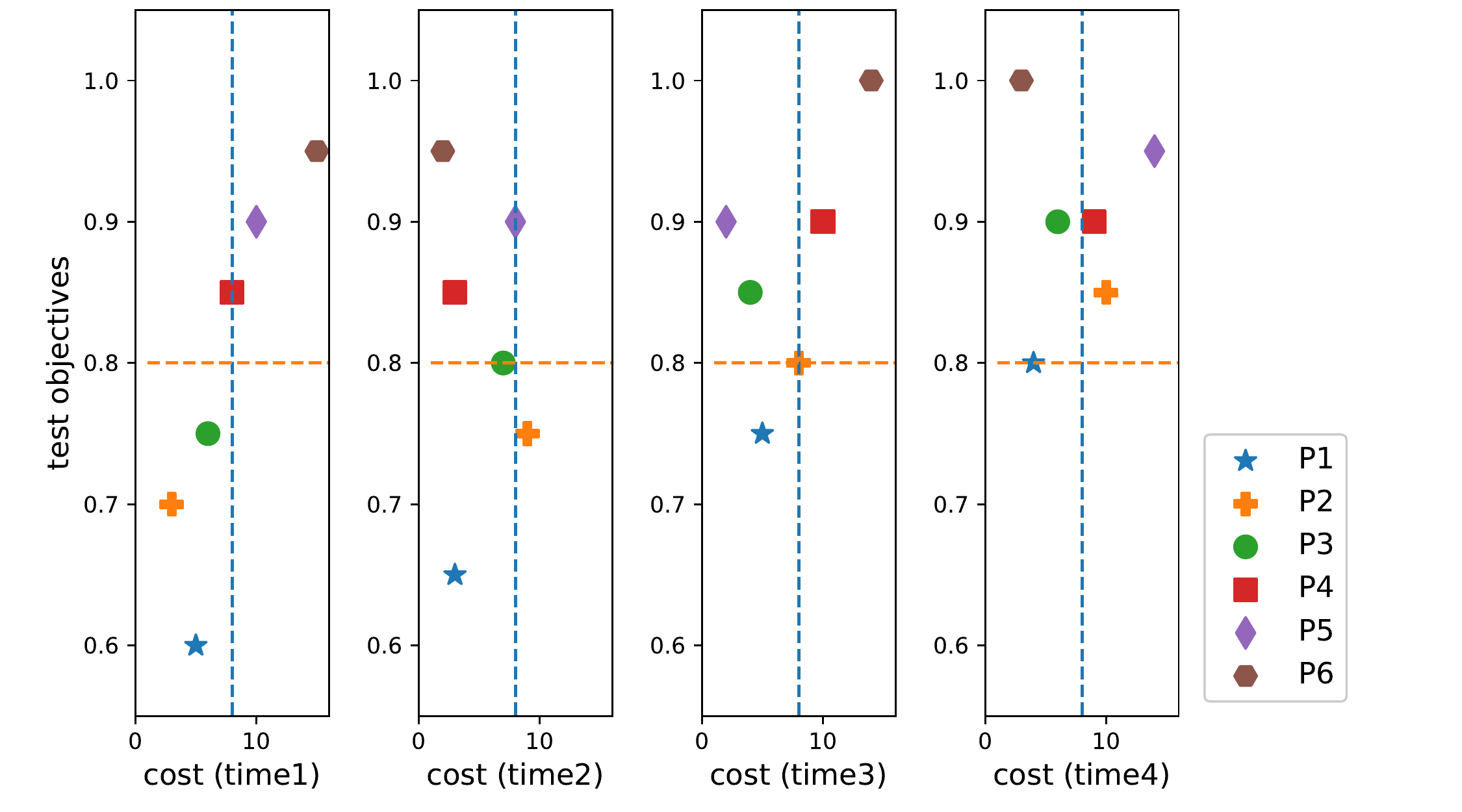}
\caption{Decision support for trade-off analysis (RQ4)}
\label{fig:trade-off}
\end{figure}

Figure \ref{fig:trade-off} demonstrates 4 trade-off analysis examples across 6 tasks (i.e. \textit{P1-P6}), generated from repeating the above analysis at four different time points  (i.e. corresponding to \textit{time1 to time4} in a sequential order). 
The y-axis denotes the next test objective to achieve, while x-axis shows the predicted required cost to achieve that objective. 

Generally speaking, the crowdtesting tasks in the right area are less cost-effective than the tasks in the left area. 
For example, at \textit{time3}, \textit{P6} is estimated to require additional 14 cost in order to achieve 90\% test objective. If the manager is facing budget constraints or trying to improve cost-effectiveness, he/she could choose to close \textit{P6} at \textit{time3}, because it is the least effective one among all tasks.
In another example, at \textit{time1}, \textit{P2} is estimated to only require 3 additional cost to reach the next objective (i.e., 70\%). This suggests the investment on 3 extra cost is highly worthwhile in increasing its quality to the next objective.  

To facilitate such kind of trade-off analysis on which task to close and when to close, we design two decision parameters as inputs from decision maker:
1) \textit{quality benchmark} which sets the minimal threshold for bug detection level, e.g. the horizontal red lines in Figure \ref{fig:trade-off}; 
2) \textit{cost benchmark} which sets the maximal threshold for test cost to achieve the next objective, e.g. the vertical blue lines in Figure \ref{fig:trade-off}. 

These two benchmarks split the tasks into four regions at each slicing time (as indicated by the four boxes in each subfigure of Figure \ref{fig:trade-off}).
Each region suggests different insights on the test sufficiency as well as cost-effectiveness for more testing, which can be used as heuristics to guide actionable decision-making at run time. More specifically:
\begin{itemize}
\item \textbf{Lower-Left (Continue)}: Tasks in this region are low hanging fruits, only requiring relatively less cost to achieve next test objective, and quality level is not acceptable yet; this indicates the most cost-effective option and testing should definitely continue. 

\item \textbf{Lower-Right (Drill down)}: Tasks here have not met the quality benchmark, so continue testing is preferred even though they require significant more cost to achieve quality objective. In addition, it likely suggests that the task is either difficult to test, or the current crowdworker participation is not sufficient. Therefore, managers would probably want to drill down in these tasks, and see if more testing guidelines or worker incentives are needed. 

\item \textbf{Upper-Left (Think twice)}: Tasks here already meet their quality benchmark, possibly reaching next higher quality level if with little additional cost investment. Managers should think twice before they take the action. 

\item \textbf{Upper-Right (Close)}: Tasks in this region require relatively more cost to reach next test objective, and current bug detection level is already high enough. This indicates that it is practical to close them considering the cost-effectiveness. 

\end{itemize}

Note that, the two benchmarks in Figure \ref{fig:trade-off} can be customized according to practical needs. 




\textbf{Answers to RQ4:} {\tool} provides practical insights to help managers make trade-off analysis on which task to close or when to close, based on two benchmark parameters and a set of decision heuristics. 



\section{Discussion}
\label{sec:discussion}

\subsection{Best CRC Estimator for Crowdtesting}
\label{sec:best_CRC}

In traditional software inspection or testing activities, \textit{MhJK}, \textit{MhCH}, and \textit{MtCH} have been recognized as the most effective estimators for total bugs \cite{briand2000comprehensive,walia2009evaluating,chun2006estimating,rong2017towards,mandala2012application,goswami2015using,vitharana2017defect}.
However, in crowdtesting, the most comprehensive estimator \textit{Mth} (see Section \ref{subsec:background_related_CRC}) outperforms the other CRC estimators. 
This is reasonable because crowdtesting is conducted by a diversified pool of crowdworkers with different levels of capability, experience, testing devices, and the nature of bugs in the software under test also vary greatly in terms of types, causes, and detection difficulty, etc. 
In such cases, \textit{Mth}, which assumes different detection probability for both bugs and crowdworkers (see Section \ref{subsec:background_related_CRC}), supposes to be the most suitable estimator for crowdtesting.

\subsection{Necessity for More Time-Sensitive Analytics in Crowdtesting Decision Support}
\label{sec:Time-sensitive decision making}

As discussed earlier in the background and motivational pilot study (Section \ref{subsec:background_observations}), challenges associated with crowdtesting management mainly lie in two aspects: uncertainty in crowdworker's performance and lack of visibility into crowdtesting progress. We believe there is an increasing need for introducing more time-sensitive analytics to support better decision making to fully realize the potential benefits of crowdtesting. 

Compared with the two baselines, {\tool} provides additional visibility into the testing progress and insights for effective task management. In particular, during the later stage of crowdtesting process, the performance of {\tool} is significantly better than the baselines (see Table \ref{tab:future_cost}).   

As discussed in answering RQ4, {\tool} can generate time-based information revealing dynamic crowdtesting progress and provide practical guidelines to help managers make trade-off analysis on which task to close or when to close, based on a set of decision heuristics. 

This suggests a significant portion of crowdtesting cost can be saved through employing effective decision support approaches such as {\tool}. This is extremely encouraging and we look forward to more discussion and innovative decision support techniques in this direction.

\subsection{Threats to Validity}
\label{subsec:dis_threats}

The external threats concern the generality of this study.
Firstly, our experiment data consists of 218 crowdtesting tasks collected from one of the Chinese largest crowdtesting platforms.
We can not assume that the results of our study could generalize beyond this environment in which it was conducted.
However, the diversity of tasks and size of dataset relatively reduce this risk. 
Secondly, our designed methods are largely dependent on the report's attributes (i.e., whether it contains a bug; and whether it is the duplicates of previous ones) assigned by the manager.
This is addressed to some extent due to the fact that we collected the data after the crowdtesting tasks were closed, and they have no knowledge about this study to artificially modify their assignment.

Internal validity of this study mainly questions the baselines. 
As there is no existing methods for managing crowdtesting tasks, we choose one commonly-used method for managing software quality, and one method based on empirical observations of crowdtesting, as the baselines to demonstrate the advantage of our proposed {\tool}.

Construct validity of this study mainly concerns the experimental setup for determining the parameter value. 
We use the most frequent tuned optimal parameter values, which can allleviate the randomness, to examine the performance of our proposed {\tool}.

\section{Related Work}
\label{sec:related}


\subsection{Crowdtesting}
\label{subsec:related_crowdtesting}

Crowdtesting has been applied to facilitate many testing tasks, e.g.,
test case generation~\cite{chen2012puzzle}, usability testing~\cite{gomide2014affective}, software performance analysis~\cite{musson2013leveraing}, software bug detection and reproduction~\cite{maria2016reproducing}.
These studies leverage crowdtesting to solve the problems in traditional testing activities,
some other approaches focus on solving the new encountered problems in crowdtesting.

Feng et al. \cite{feng2015test,feng2016multi} proposed approaches to prioritize test reports in crowdtesting.
They designed strategies to dynamically select the most risky and diversified test report for inspection in each iteration.
Jiang et al. \cite{jiang2018fuzzy} proposed the test report fuzzy clustering framework to reduce the number of inspected test reports.
Wang et al.~\cite{wang2016towards,wang2016local,wang2017domain} proposed approaches to automatically classify crowdtesting reports.
Cui et al. \cite{cui2017who,cui2017multi} and Xie et al.~\cite{xie2017cocoon} proposed crowdworker selection approaches to recommend appropriate crowdworkers for specific crowdtesting tasks.

In this work, we focus on the automated decision support for crowdtesting management, which is valuable to improve the cost-effectiveness of crowdtesting and not explored before.

\subsection{Software Quality Management}
\label{subsec:related_inspection}

Many existing approaches proposed risk-driven or value-based analysis to prioritize or select test cases \cite{wang2017qtep,shi2015comparing,harman2015empirical,saha2015information,henard2016comparing,panichella2015improving}, so as to improve the cost-effectiveness of testing.
However, none of these is applicable to the emerging crowd testing paradigm where managers typically have no control over online crowdworkers's dynamic behavior and uncertain performance.

There are also existing researches focusing on defect prediction and effort estimation \cite{menzies2007data,nam2017heterogeneous,agrawal2017better,kocaguneli2012value,singh2015systematic}. 
The core part of these approaches is the extraction of features from the source code, or software repositories. 
However, in crowdtesting, the platform can neither obtain the source code of these apps, nor involve in the software development process of these apps.

Many existing approaches focused on applying over-sampling and under-sampling techniques to alleviate the data imbalance problem in predictions \cite{wang2016automatically,tan2015online,gao2014use}.
However, what we faced is not the data imbalance problem, but the dynamic and uncertain bug arrival data.
This is why we employed incremental sampling in this study.

Several researches focused on studying the time series models for measuring software reliability \cite{Zeitler91,Bai05,Fenton08,Wiper12,Yang10,Moura11,iqbal2013software}.
Among these, ARIMA is the most promising model for mapping system failures over time.
It has been applied in estimating software failures \cite{amin2013approach}, predicting the evolution in maintenance phase of system project \cite{chong1988analyzing}, predicting the monthly number of changes of a software project \cite{kemerer1999empirical}, modeling time series changes of software \cite{herraiz2007forecasting}, etc.
This paper used ARIMA in modeling the bug arrival dynamics in crowdtesting and estimating future trend.

Another body of previous researches aimed at optimizing software inspection by predicting the total and remaining number of bugs.
Eick et al. \cite{eick1992estimating} reported the first work on employing capture-recapture models in software inspections to estimate the number of faults remaining in requirements and design artifacts.
Following that, several researches focused on evaluating the influence of number of inspectors, the number of actual defects, the dependency within inspectors, the learning style of individual inspectors, on the capture-recapture estimators' accuracy \cite{briand2000comprehensive,walia2009evaluating,chun2006estimating,rong2017towards,mandala2012application,goswami2015using,vitharana2017defect}.
The aforementioned approaches are based on different types of capture-recapture models, and results turned out \textit{MhJK}, \textit{MhCH}, and \textit{MtCH} are the most effective estimators.
We have reused all these estimators and experimentally evaluated them in crowdtesting.

\section{Conclusion}
\label{sec:conclusion}

Benefits of crowdtesting have largely been attributed to its potential to get test done faster and cheaper. Motivated by the empirical observations from an industry crowdtesting platform, this study aimed at developing automated decision support to address management blindness and achieve additional cost-effectiveness. 

The proposed {\tool} employ incremental sampling technique to address the dynamic, parallel characteristics of bug arrival data, and integrate two classical prediction models, i.e. CRC and ARIMA, to raise managers' awareness of testing progress through two indicators (i.e., total number of bugs, required cost to achieve certain test objectives).
Based on the indicators, {\tool} can be used to automate the task closing and semi-automate trade-off decisions. 
Results show that decision automation using {\tool} can largely improve the cost-effectiveness of crowdtesting. 
Specifically, a median of 100\% bugs can be detected with 30\% cost reduction. 


It should be pointed out that the presented material is just the starting point of the work in progress. 
We are collaborating with {\company} and begin to deploy {\tool} online. 
Future work includes conducting further evaluation on a broader scope of datasets, incorporating more real-world crowdtesting application scenarios, conducting more evaluational experiments in industry settings, and improving the usability of {\tool} based on evaluation feedback.


\bibliographystyle{plain}

\begin{thebibliography}{10}
\providecommand{\url}[1]{#1}
\csname url@samestyle\endcsname
\providecommand{\newblock}{\relax}
\providecommand{\bibinfo}[2]{#2}
\providecommand{\BIBentrySTDinterwordspacing}{\spaceskip=0pt\relax}
\providecommand{\BIBentryALTinterwordstretchfactor}{4}
\providecommand{\BIBentryALTinterwordspacing}{\spaceskip=\fontdimen2\font plus
\BIBentryALTinterwordstretchfactor\fontdimen3\font minus
  \fontdimen4\font\relax}
\providecommand{\BIBforeignlanguage}[2]{{%
\expandafter\ifx\csname l@#1\endcsname\relax
\typeout{** WARNING: IEEEtran.bst: No hyphenation pattern has been}%
\typeout{** loaded for the language `#1'. Using the pattern for}%
\typeout{** the default language instead.}%
\else
\language=\csname l@#1\endcsname
\fi
#2}}
\providecommand{\BIBdecl}{\relax}
\BIBdecl

\bibitem{crowdsourced_ESEM2016}
J.~Wang, Q.~Cui, Q.~Wang, and S.~Wang, ``Towards effectively test report
  classification to assist crowdsourced testing,'' in \emph{ESEM '16}, 2016,
  pp. 6:1--6:10.

\bibitem{crowd_test_report_prioritization}
Y.~Feng, Z.~Chen, J.~A. Jones, C.~Fang, and B.~Xu, ``Test report prioritization
  to assist crowdsourced testing,'' in \emph{FSE '15}, pp. 225--236.

\bibitem{junjie_ase2016}
J.~Wang, S.~Wang, Q.~Cui, and Q.~Wang, ``Local-based active classification of
  test report to assist crowdsourced testing,'' in \emph{ASE '16}, 2016, pp.
  190--201.

\bibitem{crowd_report_prioritization_ase2016}
Y.~Feng, J.~A. Jones, Z.~Chen, and C.~Fang, ``Multi-objective test report
  prioritization using image understanding,'' in \emph{ASE '16}, 2016, pp.
  202--213.

\bibitem{junjie_icse2017}
J.~Wang, Q.~Cui, S.~Wang, and Q.~Wang, ``Domain adaptation for test report
  classification in crowdsourced testing,'' in \emph{ICSE-SEIP'17}, pp. 83--92.

\bibitem{duplicate_topic_model}
A.~T. Nguyen, T.~T. Nguyen, T.~N. Nguyen, D.~Lo, and C.~Sun, ``Duplicate bug
  report detection with a combination of information retrieval and topic
  modeling,'' in \emph{ASE'12}, pp. 70--79.

\bibitem{duplicate_embedding}
X.~Yang, D.~Lo, X.~Xia, L.~Bao, and J.~Sun, ``Combining word embedding with
  information retrieval to recommend similar bug reports,'' in \emph{ISSRE'16},
  2016, pp. 127--137.

\bibitem{sun2010discriminative}
C.~Sun, D.~Lo, X.~Wang, J.~Jiang, and S.~Khoo, ``A discriminative model
  approach for accurate duplicate bug report retrieval,'' in \emph{ICSE'10},
  pp. 45--54.

\bibitem{runeson2007detection}
P.~Runeson, M.~Alexandersson, and O.~Nyholm, ``Detection of duplicate defect
  reports using natural language processing,'' in \emph{ICSE'07}, pp. 499--510.

\bibitem{tian2012improved}
Y.~Tian, C.~Sun, and D.~Lo, ``Improved duplicate bug report identification,''
  in \emph{CSMR'12}, pp. 385--390.

\bibitem{wang2008approach}
X.~Wang, L.~Zhang, T.~Xie, J.~Anvik, and J.~Sun, ``An approach to detecting
  duplicate bug reports using natural language and execution information,'' in
  \emph{ICSE'08}, pp. 461--470.

\bibitem{hindle2016contextual}
A.~Hindle, A.~Alipour, and E.~Stroulia, ``A contextual approach towards more
  accurate duplicate bug report detection and ranking,'' \emph{Empirical
  Software Engineering}, vol.~21, no.~2, pp. 368--410, 2016.

\bibitem{banerjee2013fusion}
S.~Banerjee, Z.~Syed, J.~Helmick, and B.~Cukic, ``A fusion approach for
  classifying duplicate problem reports,'' in \emph{ISSRE'13}, pp. 208--217.

\bibitem{klein2014new}
N.~Klein, C.~S. Corley, and N.~A. Kraft, ``New features for duplicate bug
  detection,'' in \emph{MSR'14}, pp. 324--327.

\bibitem{alipour2013contextual}
A.~Alipour, A.~Hindle, and E.~Stroulia, ``A contextual approach towards more
  accurate duplicate bug report detection,'' in \emph{MSR'13}, pp. 183--192.

\bibitem{duplicate_SANER_empirical}
H.~Rocha, M.~T. Valente, H.~Marques-Neto, and G.~C. Murphy, ``An empirical
  study on recommendations of similar bugs,'' in \emph{SANER'16}, vol.~1, 2016,
  pp. 46--56.

\bibitem{crowd_test_puzzle_test_case_generation}
N.~Chen and S.~Kim, ``Puzzle-based automatic testing: Bringing humans into the
  loop by solving puzzles,'' in \emph{ASE '12}, pp. 140--149.

\bibitem{crowd_test_usability_test}
V.~H.~M. Gomide, P.~A. Valle, J.~O. Ferreira, J.~R.~G. Barbosa, A.~F. da~Rocha,
  and T.~M.~G. d.~A.~Barbosa, ``Affective crowdsourcing applied to usability
  testing,'' \emph{Int. J. of Computer Science and Information Technologies},
  vol.~5, no.~1, pp. 575--579, 2014.

\bibitem{crowd_test_performance}
R.~Musson, J.~Richards, D.~Fisher, C.~Bird, B.~Bussone, and S.~Ganguly,
  ``Leveraging the crowd: How 48,000 users helped improve lync performance,''
  \emph{IEEE Software}, vol.~30, no.~4, pp. 38--45, 2013.

\bibitem{mobile16}
M.~G., R.~R., B.~A., and L.~S., ``Reproducing context-sensitive crashes of
  mobile apps using crowdsourced monitoring,'' in \emph{MOBILESoft '16}, pp.
  88--99.

\bibitem{cuiqiang_seke2017}
Q.~Cui, S.~Wang, J.~Wang, Y.~Hu, Q.~Wang, and M.~Li, ``Multi-objective crowd
  worker selection in crowdsourced testing,'' in \emph{SEKE'17}, pp. 1--6.

\bibitem{cuiqiang_compsac2017}
Q.~Cui, J.~Wang, G.~Yang, M.~Xie, Q.~Wang, and M.~Li, ``Who should be selected
  to perform a task in crowdsourced testing?'' in \emph{COMPSAC'17}, pp.
  75--84.

\bibitem{sun2011towards}
C.~Sun, D.~Lo, S.~Khoo, and J.~Jiang, ``Towards more accurate retrieval of
  duplicate bug reports,'' in \emph{ASE'11}, pp. 253--262.

\bibitem{data_ming}
I.~H. Witten and E.~Frank, \emph{Data Mining: Practical machine learning tools
  and techniques}.\hskip 1em plus 0.5em minus 0.4em\relax Morgan Kaufmann,
  2005.

\bibitem{image_role_lifeifei}
X.~Chen and C.~Lawrence~Zitnick, ``Mind's eye: A recurrent visual
  representation for image caption generation,'' in \emph{CVPR'15}, June 2015.

\bibitem{image_role}
O.~Vinyals, A.~Toshev, S.~Bengio, and D.~Erhan, ``Show and tell: Lessons
  learned from the 2015 mscoco image captioning challenge,'' \emph{IEEE
  Transactions on Pattern Analysis and Machine Intelligence}, vol.~39, no.~4,
  pp. 652--663, April 2017.

\bibitem{image_structure_feature}
A.~Oliva and A.~Torralba, ``Modeling the shape of the scene: A holistic
  representation of the spatial envelope,'' \emph{Int. J. Comput. Vision},
  vol.~42, no.~3, pp. 145--175, 2001.

\bibitem{image_color_feature}
B.~S. Manjunath, J.~R. Ohm, V.~V. Vasudevan, and A.~Yamada, ``Color and texture
  descriptors,'' \emph{IEEE Trans. Cir. and Sys. for Video Technol.}, vol.~11,
  no.~6, pp. 703--715, 2001.

\bibitem{embedding_model}
T.~Mikolov, I.~Sutskever, K.~Chen, G.~Corrado, and J.~Dean, ``Distributed
  representations of words and phrases and their compositionality,'' in
  \emph{NIPS'13}, 2013, pp. 3111--3119.

\bibitem{embedding_model_Bengio}
Y.~Bengio, R.~Ducharme, P.~Vincent, and C.~Janvin, ``A neural probabilistic
  language model,'' \emph{The Journal of Machine Learning Research}, vol.~3,
  pp. 1137--1155, 2003.

\bibitem{duplicate_linkable_knowledge_stackoverflow}
B.~Xu, D.~Ye, Z.~Xing, X.~Xia, G.~Chen, and S.~Li, ``Predicting semantically
  linkable knowledge in developer online forums via convolutional neural
  network,'' in \emph{ASE'16}, 2016, pp. 51--62.

\end{thebibliography}


\begin{thebibliography}{10}
\providecommand{\url}[1]{#1}
\csname url@samestyle\endcsname
\providecommand{\newblock}{\relax}
\providecommand{\bibinfo}[2]{#2}
\providecommand{\BIBentrySTDinterwordspacing}{\spaceskip=0pt\relax}
\providecommand{\BIBentryALTinterwordstretchfactor}{4}
\providecommand{\BIBentryALTinterwordspacing}{\spaceskip=\fontdimen2\font plus
\BIBentryALTinterwordstretchfactor\fontdimen3\font minus
  \fontdimen4\font\relax}
\providecommand{\BIBforeignlanguage}[2]{{%
\expandafter\ifx\csname l@#1\endcsname\relax
\typeout{** WARNING: IEEEtran.bst: No hyphenation pattern has been}%
\typeout{** loaded for the language `#1'. Using the pattern for}%
\typeout{** the default language instead.}%
\else
\language=\csname l@#1\endcsname
\fi
#2}}
\providecommand{\BIBdecl}{\relax}
\BIBdecl

\bibitem{mao2017survey}
K.~Mao, L.~Capra, M.~Harman, and Y.~Jia, ``A survey of the use of crowdsourcing
  in software engineering,'' \emph{Journal of Systems and Software}, vol. 126,
  pp. 57--84, 2017.

\bibitem{chen2018research}
X.~Zhang, Y.~Feng, D.~Liu, Z.~Chen, and B.~Xu, ``Research progress of
  crowdsourced software testing,'' \emph{Journal of Software}, vol. 29(1), pp.
  69--88, 2018.

\bibitem{link_crowdtest}
\url{http://www.softwaretestinghelp.com/crowdsourced-testing-companies/}, 2018.

\bibitem{wang2017domain}
J.~Wang, Q.~Cui, S.~Wang, and Q.~Wang, ``Domain adaptation for test report
  classification in crowdsourced testing,'' in \emph{Proceedings of the 39th
  International Conference on Software Engineering: Software Engineering in
  Practice Track}.\hskip 1em plus 0.5em minus 0.4em\relax IEEE Press, 2017, pp.
  83--92.

\bibitem{cui2017who}
Q.~Cui, J.~Wang, G.~Yang, M.~Xie, Q.~Wang, and M.~Li, ``Who should be selected
  to perform a task in crowdsourced testing?'' in \emph{2017 IEEE 41st Annual
  Computer Software and Applications Conference}, vol.~1.\hskip 1em plus 0.5em
  minus 0.4em\relax IEEE, 2017, pp. 75--84.

\bibitem{myers2011art}
G.~J. Myers, C.~Sandler, and T.~Badgett, \emph{The art of software
  testing}.\hskip 1em plus 0.5em minus 0.4em\relax John Wiley \& Sons, 2011.

\bibitem{lewis2016software}
W.~E. Lewis, \emph{Software testing and continuous quality improvement}.\hskip
  1em plus 0.5em minus 0.4em\relax CRC press, 2016.

\bibitem{garg2011stop}
M.~Garg, R.~Lai, and S.~J. Huang, ``When to stop testing: a study from the
  perspective of software reliability models,'' \emph{IET software}, vol.~5,
  no.~3, pp. 263--273, 2011.

\bibitem{iqbal2013software}
J.~Iqbal, N.~Ahmad, and S.~Quadri, ``A software reliability growth model with
  two types of learning,'' in \emph{2013 International Conference on Machine
  Intelligence and Research Advancement}.\hskip 1em plus 0.5em minus
  0.4em\relax IEEE, 2013, pp. 498--503.

\bibitem{wang2017qtep}
S.~Wang, J.~Nam, and L.~Tan, ``Q{TEP}: quality-aware test case
  prioritization,'' in \emph{Proceedings of the 2017 11th Joint Meeting on
  Foundations of Software Engineering}.\hskip 1em plus 0.5em minus 0.4em\relax
  ACM, 2017, pp. 523--534.

\bibitem{shi2015comparing}
A.~Shi, T.~Yung, A.~Gyori, and D.~Marinov, ``Comparing and combining test-suite
  reduction and regression test selection,'' in \emph{Proceedings of the 2015
  10th Joint Meeting on Foundations of Software Engineering}.\hskip 1em plus
  0.5em minus 0.4em\relax ACM, 2015, pp. 237--247.

\bibitem{harman2015empirical}
M.~G. Epitropakis, S.~Yoo, M.~Harman, and E.~K. Burke, ``Empirical evaluation
  of pareto efficient multi-objective regression test case prioritisation,'' in
  \emph{Proceedings of the 2015 International Symposium on Software Testing and
  Analysis}.\hskip 1em plus 0.5em minus 0.4em\relax ACM, 2015, pp. 234--245.

\bibitem{saha2015information}
R.~K. Saha, L.~Zhang, S.~Khurshid, and D.~E. Perry, ``An information retrieval
  approach for regression test prioritization based on program changes,'' in
  \emph{2015 IEEE/ACM 37th IEEE International Conference on Software
  Engineering}, vol.~1.\hskip 1em plus 0.5em minus 0.4em\relax IEEE, 2015, pp.
  268--279.

\bibitem{henard2016comparing}
C.~Henard, M.~Papadakis, M.~Harman, Y.~Jia, and Y.~Le~Traon, ``Comparing
  white-box and black-box test prioritization,'' in \emph{2016 IEEE/ACM 38th
  International Conference on Software Engineering}.\hskip 1em plus 0.5em minus
  0.4em\relax IEEE, 2016, pp. 523--534.

\bibitem{rong2017towards}
G.~Rong, B.~Liu, H.~Zhang, Q.~Zhang, and D.~Shao, ``Towards confidence with
  capture-recapture estimation: An exploratory study of dependence within
  inspections,'' in \emph{Proceedings of the 21st International Conference on
  Evaluation and Assessment in Software Engineering}, 2017, pp. 242--251.

\bibitem{liu2015adoption}
G.~Liu, G.~Rong, H.~Zhang, and Q.~Shan, ``The adoption of capture-recapture in
  software engineering: a systematic literature review,'' in \emph{Proceedings
  of the 19th International Conference on Evaluation and Assessment in Software
  Engineering}.\hskip 1em plus 0.5em minus 0.4em\relax ACM, 2015, p.~15.

\bibitem{chun2006estimating}
Y.~H.Chun, ``Estimating the number of undetected software errors via the
  correlated capture-recapture model,'' \emph{European Journal of Operational
  Research}, vol. 175, no.~2, pp. 1180 -- 1192, 2006.

\bibitem{mandala2012application}
N.~R. Mandala, G.~S. Walia, J.~C. Carver, and N.~Nagappan, ``Application of
  kusumoto cost-metric to evaluate the cost effectiveness of software
  inspections,'' in \emph{Proceedings of the ACM-IEEE International Symposium
  on Empirical Software Engineering and Measurement}, 2012, pp. 221--230.

\bibitem{amin2013approach}
A.~Amin, L.~Grunske, and A.~Colman, ``An approach to software reliability
  prediction based on time series modeling,'' \emph{Journal of Systems and
  Software}, vol.~86, no.~7, pp. 1923--1932, 2013.

\bibitem{chong1988analyzing}
C.~Chong Hok~Yuen, ``On analyzing maintenance process data at the global and
  the detailed levels: A case study,'' in \emph{Proceedings of the IEEE
  Conference on Software Maintenance}, 1988, pp. 248--255.

\bibitem{kemerer1999empirical}
C.~F. Kemerer and S.~Slaughter, ``An empirical approach to studying software
  evolution,'' \emph{IEEE Transactions on Software Engineering}, vol.~25,
  no.~4, pp. 493--509, 1999.

\bibitem{herraiz2007forecasting}
I.~Herraiz, J.~M. Gonzalez-Barahona, and G.~Robles, ``Forecasting the number of
  changes in eclipse using time series analysis,'' in \emph{Fourth
  International Workshop on Mining Software Repositories}.\hskip 1em plus 0.5em
  minus 0.4em\relax IEEE, 2007, pp. 32--32.

\bibitem{wang2016towards}
J.~Wang, Q.~Cui, Q.~Wang, and S.~Wang, ``Towards effectively test report
  classification to assist crowdsourced testing,'' in \emph{Proceedings of the
  10th ACM/IEEE International Symposium on Empirical Software Engineering and
  Measurement}.\hskip 1em plus 0.5em minus 0.4em\relax ACM, 2016, p.~6.

\bibitem{incrementalsampling}
L.~Mora-Applegate and M.~Malinowski, ``Incremental sampling methodology,''
  Interstate Technology and Regulatory Council (ITRC), Tech. Rep., 2012.

\bibitem{Mthlee1996estimating}
S.-M. Lee, ``Estimating population size for capture-recapture data when capture
  probabilities vary by time, behavior and individual animal,''
  \emph{Communications in Statistics-Simulation and Computation}, vol.~25,
  no.~2, pp. 431--457, 1996.

\bibitem{M0laplace1783naissances}
P.~S. Laplace, ``Sur les naissances, les mariages et les morts,''
  \emph{Histaire de I'Academie Royale des Sciences}, p. 693, 1783.

\bibitem{MtCHchao1987estimating}
A.~Chao, ``Estimating the population size for capture-recapture data with
  unequal catchability,'' \emph{Biometrics}, pp. 783--791, 1987.

\bibitem{MhCHchao1988estimating}
------, ``Estimating animal abundance with capture frequency data,'' \emph{The
  Journal of Wildlife Management}, pp. 295--300, 1988.

\bibitem{MhJKburnham1978estimation}
K.~P. Burnham and W.~S. Overton, ``Estimation of the size of a closed
  population when capture probabilities vary among animals,''
  \emph{Biometrika}, vol.~65, no.~3, pp. 625--633, 1978.

\bibitem{tan2015online}
M.~Tan, L.~Tan, S.~Dara, and C.~Mayeux, ``Online defect prediction for
  imbalanced data,'' in \emph{Proceedings of the 37th International Conference
  on Software Engineering-Volume 2}.\hskip 1em plus 0.5em minus 0.4em\relax
  IEEE Press, 2015, pp. 99--108.

\bibitem{nam2017heterogeneous}
J.~Nam, W.~Fu, S.~Kim, T.~Menzies, and L.~Tan, ``Heterogeneous defect
  prediction,'' \emph{IEEE Transactions on Software Engineering}, 2017.

\bibitem{kan2002metrics}
S.~H. Kan, \emph{Metrics and models in software quality engineering}.\hskip 1em
  plus 0.5em minus 0.4em\relax Addison-Wesley Longman Publishing Co., Inc.,
  2002.

\bibitem{briand2000comprehensive}
L.~C. Briand, K.~E. Emam, B.~G. Freimut, and O.~Laitenberger, ``A comprehensive
  evaluation of capture-recapture models for estimating software defect
  content,'' \emph{IEEE Transactions on Software Engineering}, vol.~26, no.~6,
  pp. 518--540, 2000.

\bibitem{walia2009evaluating}
G.~S. Walia and J.~C. Carver, ``Evaluating the effect of the number of
  naturally occurring faults on the estimates produced by capture-recapture
  models,'' in \emph{2009 International Conference on Software Testing
  Verification and Validation}, 2009, pp. 210--219.

\bibitem{goswami2015using}
A.~Goswami, G.~Walia, and A.~Singh, ``Using learning styles of software
  professionals to improve their inspection team performance,''
  \emph{International Journal of Software Engineering and Knowledge
  Engineering}, vol.~25, no. 09-10, pp. 1721--1726, 2015.

\bibitem{vitharana2017defect}
P.~Vitharana, ``Defect propagation at the project-level: results and a post-hoc
  analysis on inspection efficiency,'' \emph{Empirical Software Engineering},
  vol.~22, no.~1, pp. 57--79, 2017.

\bibitem{chen2012puzzle}
N.~Chen and S.~Kim, ``Puzzle-based automatic testing: Bringing humans into the
  loop by solving puzzles,'' in \emph{Proceedings of the 27th IEEE/ACM
  International Conference on Automated Software Engineering}.\hskip 1em plus
  0.5em minus 0.4em\relax ACM, 2012, pp. 140--149.

\bibitem{gomide2014affective}
V.~H.~M. Gomide, P.~A. Valle, J.~O. Ferreira, J.~R.~G. Barbosa, A.~F. da~Rocha,
  and T.~M.~G. d.~A.~Barbosa, ``Affective crowdsourcing applied to usability
  testing,'' \emph{Int. J. of Computer Science and Information Technologies},
  vol.~5, no.~1, pp. 575--579, 2014.

\bibitem{musson2013leveraing}
R.~Musson, J.~Richards, D.~Fisher, C.~Bird, B.~Bussone, and S.~Ganguly,
  ``Leveraging the crowd: How 48,000 users helped improve lync performance,''
  \emph{IEEE Software}, vol.~30, no.~4, pp. 38--45, 2013.

\bibitem{maria2016reproducing}
M.~G{\'o}mez, R.~Rouvoy, B.~Adams, and L.~Seinturier, ``Reproducing
  context-sensitive crashes of mobile apps using crowdsourced monitoring,'' in
  \emph{2016 IEEE/ACM International Conference on Mobile Software Engineering
  and Systems}.\hskip 1em plus 0.5em minus 0.4em\relax IEEE, 2016, pp. 88--99.

\bibitem{feng2015test}
Y.~Feng, Z.~Chen, J.~A. Jones, C.~Fang, and B.~Xu, ``Test report prioritization
  to assist crowdsourced testing.'' in \emph{Proceedings of the 2015 10th Joint
  Meeting on Foundations of Software Engineering}, 2015, pp. 225--236.

\bibitem{feng2016multi}
Y.~Feng, J.~A. Jones, Z.~Chen, and C.~Fang, ``Multi-objective test report
  prioritization using image understanding,'' in \emph{2016 31st IEEE/ACM
  International Conference on Automated Software Engineering}.\hskip 1em plus
  0.5em minus 0.4em\relax IEEE, 2016, pp. 202--213.

\bibitem{jiang2018fuzzy}
H.~Jiang, X.~Chen, T.~He, Z.~Chen, and X.~Li, ``Fuzzy clustering of
  crowdsourced test reports for apps,'' \emph{ACM Transactions on Internet
  Technology}, vol.~18, no.~2, p.~18, 2018.

\bibitem{wang2016local}
J.~Wang, S.~Wang, Q.~Cui, and Q.~Wang, ``Local-based active classification of
  test report to assist crowdsourced testing,'' in \emph{2016 31st
  International Conference on Automated Software Engineering}.\hskip 1em plus
  0.5em minus 0.4em\relax IEEE, 2016, pp. 190--201.

\bibitem{cui2017multi}
Q.~Cui, S.~Wang, J.~Wang, Y.~Hu, Q.~Wang, and M.~Li, ``Multi-objective crowd
  worker selection in crowdsourced testing,'' in \emph{29th International
  Conference on Software Engineering and Knowledge Engineering}, 2017, pp.
  218--223.

\bibitem{xie2017cocoon}
M.~Xie, Q.~Wang, G.~Yang, and M.~Li, ``Cocoon: Crowdsourced testing quality
  maximization under context coverage constraint,'' in \emph{2017 IEEE 28th
  International Symposium on Software Reliability Engineering}.\hskip 1em plus
  0.5em minus 0.4em\relax IEEE, 2017, pp. 316--327.

\bibitem{panichella2015improving}
A.~Panichella, R.~Oliveto, M.~Di~Penta, and A.~De~Lucia, ``Improving
  multi-objective test case selection by injecting diversity in genetic
  algorithms,'' \emph{IEEE Transactions on Software Engineering}, vol.~41,
  no.~4, pp. 358--383, 2015.

\bibitem{menzies2007data}
T.~Menzies, J.~Greenwald, and A.~Frank, ``Data mining static code attributes to
  learn defect predictors,'' \emph{IEEE transactions on software engineering},
  vol.~33, no.~1, pp. 2--13, 2007.

\bibitem{agrawal2017better}
A.~Agrawal and T.~Menzies, ````better data'' is better than ``better data
  miners'' (benefits of tuning smote for defect prediction),'' in
  \emph{Proceedings of the 40th International Conference on Software
  engineering}, 2018.

\bibitem{kocaguneli2012value}
E.~Kocaguneli, T.~Menzies, and J.~W. Keung, ``On the value of ensemble effort
  estimation,'' \emph{IEEE Transactions on Software Engineering}, vol.~38,
  no.~6, pp. 1403--1416, 2012.

\bibitem{singh2015systematic}
P.~K. Singh, D.~Agarwal, and A.~Gupta, ``A systematic review on software defect
  prediction,'' in \emph{2015 2nd International Conference on Computing for
  Sustainable Global Development}.\hskip 1em plus 0.5em minus 0.4em\relax IEEE,
  2015, pp. 1793--1797.

\bibitem{wang2016automatically}
S.~Wang, T.~Liu, and L.~Tan, ``Automatically learning semantic features for
  defect prediction,'' in \emph{Proceedings of the 38th International
  Conference on Software Engineering}.\hskip 1em plus 0.5em minus 0.4em\relax
  ACM, 2016, pp. 297--308.

\bibitem{gao2014use}
K.~GAO, T.~M. KHOSHGOFTAAR, and R.~WALD, ``The use of under-and oversampling
  within ensemble feature selection and classification for software quality
  prediction,'' \emph{International Journal of Reliability, Quality and Safety
  Engineering}, vol.~21, no.~01, p. 1450004, 2014.

\bibitem{Zeitler91}
D.~Zeitler, ``Realistic assumptions for software reliability models,'' in
  \emph{Software Reliability Engineering, 1991. Proceedings., 1991
  International Symposium on}.\hskip 1em plus 0.5em minus 0.4em\relax IEEE,
  1991, pp. 67--74.

\bibitem{Bai05}
C.~Bai, Q.~Hu, M.~Xie, and S.~H. Ng, ``Software failure prediction based on a
  markov bayesian network model,'' \emph{Journal of Systems and Software},
  vol.~74, no.~3, pp. 275--282, 2005.

\bibitem{Fenton08}
N.~Fenton, M.~Neil, and D.~Marquez, ``Using bayesian networks to predict
  software defects and reliability,'' \emph{Proceedings of the Institution of
  Mechanical Engineers, Part O: Journal of Risk and Reliability}, vol. 222,
  no.~4, pp. 701--712, 2008.

\bibitem{Wiper12}
M.~Wiper, A.~Palacios, and J.~Mar{\'\i}n, ``Bayesian software reliability
  prediction using software metrics information,'' \emph{Quality Technology \&
  Quantitative Management}, vol.~9, no.~1, pp. 35--44, 2012.

\bibitem{Yang10}
B.~Yang, X.~Li, M.~Xie, and F.~Tan, ``A generic data-driven software
  reliability model with model mining technique,'' \emph{Reliability
  Engineering \& System Safety}, vol.~95, no.~6, pp. 671--678, 2010.

\bibitem{Moura11}
M.~das Chagas~Moura, E.~Zio, I.~D. Lins, and E.~Droguett, ``Failure and
  reliability prediction by support vector machines regression of time series
  data,'' \emph{Reliability Engineering \& System Safety}, vol.~96, no.~11, pp.
  1527--1534, 2011.

\bibitem{eick1992estimating}
S.~G. Eick, C.~R. Loader, M.~D. Long, L.~G. Votta, and S.~Vander~Wiel,
  ``Estimating software fault content before coding,'' in \emph{Proceedings of
  the 14th international conference on Software engineering}, 1992, pp. 59--65.

\end{thebibliography}


\end{document}